\documentclass[fleqn,usenatbib]{mnras}

\usepackage{newtxtext,newtxmath}

\usepackage{amsmath}

\usepackage[T1]{fontenc}
\usepackage{ae,aecompl}

\usepackage[normalem]{ulem}
\usepackage{graphicx}	
\usepackage{amsmath}	
\usepackage{amssymb}	
\usepackage{makecell}
\graphicspath{
{./Figures/}
}

\usepackage{orcidlink}

\newcommand{\f}{\frac}
\newcommand{\tx}{\text}

\title[Phase Resolved \textit{JWST} MIRI Spectra of WASP-43b]{Simultaneous retrieval of orbital phase resolved \textit{JWST}/MIRI emission spectra of the hot Jupiter WASP-43b: evidence of water, ammonia and carbon monoxide}

\author[J. Yang et al.]{
Jingxuan Yang$^{\orcidlink{0009-0006-2395-6197}}$,$^{1}$\thanks{E-mail: jingxuanyang15@gmail.com}
Mark Hammond$^{\orcidlink{0000-0002-6893-522X}}$,$^{1}$
Anjali A. A. Piette$^{\orcidlink{0000-0002-4487-5533}}$,$^{2}$
Jasmina Blecic$^{\orcidlink{0000-0002-0769-9614}}$,$^{3,4}$
\newauthor
Taylor J. Bell$^{\orcidlink{0000-0003-4177-2149}}$,$^{5,6}$
Patrick G.J. Irwin$^{\orcidlink{0000-0002-6772-384X}}$,$^{1}$
Vivien Parmentier$^{\orcidlink{0000-0001-9521-6258}}$,$^{7}$
Shang-Min Tsai$^{\orcidlink{0000-0002-8163-4608}}$,$^{8}$
\newauthor
Joanna K. Barstow$^{\orcidlink{0000-0003-3726-5419}}$,$^{9}$
Nicolas Crouzet$^{\orcidlink{0000-0001-7866-8738}}$,$^{10}$
Laura Kreidberg$^{\orcidlink{0000-0003-0514-1147}}$,$^{11}$
Jo\~ao M. Mendon\c ca$^{\orcidlink{0000-0002-6907-4476}}$,$^{12}$
\newauthor
Jake Taylor$^{\orcidlink{0000-0003-4844-9838}}$,$^{13}$
Robin Baeyens$^{\orcidlink{0000-0001-7578-969X}}$,$^{14}$
Kazumasa Ohno$^{\orcidlink{0000-0003-3290-6758}}$,$^{15}$
Lucas Teinturier$^{\orcidlink{0000-0002-0797-5746}}$, $^{16,17}$
\newauthor
Matthew C. Nixon$^{\orcidlink{0000-0001-8236-5553}}$ $^{18}$
\\
$^{1}$Atmospheric, Oceanic and Planetary Physics, Department of Physics, University of Oxford, Oxford OX1 3PU, UK\\
$^{2}$Carnegie Institution for Science, Earth and Planets Laboratory, Washington, DC, USA\\
$^{3}$Department of Physics, New York University Abu Dhabi, Abu Dhabi, UAE\\
$^{4}$Center for Astrophysics and Space Science (CASS), New York University Abu Dhabi, Abu Dhabi, UAE\\
$^{5}$BAER Institute, NASA Ames Research Center, Moffet Field, CA 94035, USA\\
$^{6}$Space Science and Astrobiology Division, NASA Ames Research Center, Moffet Field, CA 94035, USA\\
$^{7}$Université Côte d’Azur, Observatoire de la Côte d’Azur, CNRS, Laboratoire Lagrange UMR 7293\\
$^{8}$Department of Earth and Planetary Sciences, University of California, Riverside, Geology Bldg, 900 University Ave, Riverside, CA, USA\\
$^{9}$School of Physical Sciences, The Open University, Walton Hall, Milton Keynes, MK7 6AA, UK\\
$^{10}$Leiden Observatory, Leiden University, P.O. Box 9513, 2300 RA Leiden, The Netherlands\\
$^{11}$Max Planck Institute for Astronomy, K\"onigstuhl 17, D-69117 Heidelberg, Germany\\
$^{12}$National Space Institute, Technical University of Denmark, Elektrovej, 2800 Kgs. Lyngby, Denmark\\
$^{13}$Department of Physics, University of Oxford, Parks Road, Oxford OX1 3PU, UK\\
$^{14}$Anton Pannekoek Institute for Astronomy, University of Amsterdam, Science Park 904, 1098 XH Amsterdam, The Netherlands\\
$^{15}$Division of Science, National Astronomical Observatory of Japan\\
$^{16}$LESIA, Observatoire de Paris, Université PSL, Sorbonne Université, Université Paris Cité, CNRS, 5 place Jules Janssen, 92195 Meudon, France\\
$^{17}$Laboratoire de Météorologie Dynamique, IPSL, CNRS, Sorbonne Université, Ecole Normale Supérieure, Université PSL, Ecole Polytechnique, \\ Institut Polytechnique de Paris, 75005 Paris, France\\
$^{18}$Department of Astronomy, University of Maryland, 4296 Stadium Drive, College Park, MD 20742, USA
}

\date{Accepted XXX. Received YYY; in original form ZZZ}

\pubyear{2023}

\begin{document}
\label{firstpage}
\pagerange{\pageref{firstpage}--\pageref{lastpage}}
\maketitle

\begin{abstract}
Spectroscopic phase curves of hot Jupiters measure their emission spectra at multiple orbital phases, thus enabling detailed characterisation of their atmospheres.
Precise constraints on the atmospheric composition of these exoplanets offer insights into their formation and evolution. 
We analyse four phase-resolved emission spectra of the hot Jupiter WASP-43b, generated from a phase curve observed with the MIRI/LRS onboard the \textit{JWST}, to retrieve its atmospheric properties.
Using a parametric 2D temperature model and assuming a chemically homogeneous atmosphere within the observed pressure region, we simultaneously fit the four spectra to constrain the abundances of atmospheric constituents, thereby yielding more precise constraints than previous work that analysed each spectrum independently. 
Our analysis reveals statistically significant evidence of NH$_3$ (4$\sigma$) in a hot Jupiter's emission spectra for the first time, along with evidence of H$_2$O (6.5$\sigma$), CO (3.1$\sigma$), and a non-detection of CH$_4$.
With our abundance constraints, we tentatively estimate the metallicity of WASP-43b at 0.6$-$6.5$\times$ solar and its C/O ratio at 0.6$-$0.9.
Our findings offer vital insights into the atmospheric conditions and formation history of WASP-43b by simultaneously constraining the abundances of carbon, oxygen, and nitrogen-bearing species. 
\end{abstract}

\begin{keywords}
radiative transfer -- methods: numerical -- planets and satellites: atmospheres -- planets and satellites: individual: WASP-43b.
\end{keywords}

\section{Introduction}
The discovery of hot Jupiters stands as a testament to the diverse outcomes of planetary formation processes.
Understanding the mechanisms that give rise to their close-in orbits is crucial for advancing our knowledge of planetary formation, thus necessitating detailed characterisation of their physical properties.
Fortunately, because of their large atmospheric scale heights and favourable planet-to-host-star flux ratios, hot Jupiters are excellent targets for spectroscopic atmospheric characterisation \citep[e.g.,][]{sing_continuum_2016, fisher_retrieval_2018,irwin_25d_2020,mansfield_unique_2021,ahrer_identification_2023}.
Precise constraints on hot Jupiters' atmospheric properties, such as molecular composition and thermal structure, further our knowledge of atmospheric physics under extreme conditions and grant us insight into their formation and migration pathways \citep{madhusudhan_toward_2014, mordasini_imprint_2016, madhusudhan_atmospheric_2017,dawson_origins_2018,cridland_connecting_2019, chachan_breaking_2023}.
In recent years, phase-resolved emission spectroscopy, or `spectroscopic phase curves', where the disc-integrated emission spectrum of a transiting exoplanet is observed at multiple orbital phases,
has gained attention as a powerful observation strategy 
\citep{stevenson_thermal_2014, kreidberg_global_2018, irwin_25d_2020,  mikal-evans_diurnal_2022}. 
These phase curves are valuable data sets as they allow us to probe the longitudinal variation of atmospheric properties across a planet and mitigate the degeneracy between thermal structure and molecular abundances in atmospheric retrievals \citep{feng_impact_2016, blecic_implications_2017, taylor_understanding_2020}.

In this work, we retrieve the atmospheric properties of the transiting hot Jupiter WASP-43b \citep{hellier_wasp-43b_2011} using a phase curve observed with the Mid-Infrared Instrument/Low Resolution Spectrometer (MIRI/LRS) onboard the \textit{JWST}. 
WASP-43b orbits its K7 host star in just 19.2 hours \citep{gillon_trappist_2012}, making it a prime target for phase curve observations. 
The planet is presumably tidally locked due to the proximity to its host star, allowing us to convert the orbital phase to the central longitude of the observed hemisphere. 
Using the parametric atmospheric model of \cite{yang_testing_2023}, we constrain the molecular abundances in the atmosphere of WASP-43b by simultaneously fitting four emission spectra generated at different orbital phases, with the assumption that atmospheric dynamics effectively homogenise the chemical composition in the pressure range probed by the observation \citep{cooper_dynamics_2006, agundez_pseudo_2014, drummond_observable_2018, drummond_3d_2018,  mendonca_three-dimensional_2018, venot_global_2020, baeyens_grid_2021}.
By analysing the four spectra together, we are able to boost the signal-to-noise ratio and better constrain the molecular abundances than the previous analysis that fit each spectrum separately \citep{bell_nightside_2024}.
We confirm previous detection of H$_2$O using the same data set \citep{bell_nightside_2024} at higher statistical significance and additionally find statistically significant evidence of NH$_3$ and CO.
Our work adds to a wealth of studies on WASP-43b, spanning telescopic observations \citep{czesla_x-ray_2013, blecic_spitzer_2014, kreidberg_precise_2014, stevenson_thermal_2014, stevenson_spitzer_2017, weaver_access_2020, fraine_dark_2021, scandariato_phase_2022, lesjak_retrieval_2023, murphy_lack_2023, bell_nightside_2024}, atmospheric retrievals \citep{changeat_taurex3_2020, feng_2d_2020, irwin_25d_2020, cubillos_longitudinally_2021, chubb_exoplanet_2022,  dobbs-dixon_gcm-motivated_2022,  yang_testing_2023,taylor_another_2023} and atmospheric modelling \citep{kataria_atmospheric_2015, keating_revisiting_2017, mendonca_revisiting_2018, mendonca_three-dimensional_2018, venot_global_2020, helling_mineral_2020, carone_equatorial_2020, schneider_exploring_2022, teinturier_radiative_2024}.
Our simultaneous abundance constraints on water, methane, and ammonia can enable further studies on both the atmospheric chemistry and the formation history of WASP-43b \citep{fortney_beyond_2020, line_solar_2021, ohno_nitrogen_2023, ohno_nitrogen_2023-1}.
 
This paper is structured as follows.
In the methodology section (section \ref{sec:methodology}), we first outline how the emission spectra are generated from the \textit{JWST} MIRI/LRS phase curve in \ref{sec:observation}, then introduce our parametric atmospheric model in \ref{sec:model}, our radiative transfer routine in \ref{sec:radtran}, and the set-up of our atmospheric retrievals in \ref{sec:retrieval}.
We present our results in section \ref{sec:results}, including the spectral fits in \ref{sec:spectral_fit}, the abundance constraints in \ref{sec:abundance} and the thermal structure constraints in \ref{sec:thermal_structure}.
We compare our results to past \textit{HST} and \textit{Spitzer} observations in \ref{sec:HST} and briefly discuss the implications of our results on the formation history of WASP-43b in \ref{sec:formation}.  
We explore the implications of our retrieved atmospheric properties with 1D chemical modelling in \ref{sec:1D_chem}.
We detail the limitations of the assumptions made in our analysis in section \ref{sec:assumptions} before concluding in section \ref{sec:conclusion}.

\section{Methodology}
\label{sec:methodology}

We describe the \textit{JWST} observation in \ref{sec:observation}, followed by an outline of our atmospheric retrieval procedure, which is divided into three parts. 
Firstly, we model the atmosphere of WASP-43b using the global parametric atmospheric model from \cite{yang_testing_2023} based on a radiative-equilibrium temperature-pressure profile and assuming uniform molecular abundances (\ref{sec:model}). 
Secondly, we use a correlated-k radiative transfer pipeline to generate phase-resolved disc-integrated emission spectra from a given atmospheric model (\ref{sec:radtran}).
Thirdly, we use a Bayesian parameter inference algorithm to constrain the parameters of our atmospheric model given the observed spectra and calculate the significance of the molecular detections from Bayesian evidence (\ref{sec:retrieval}).

\subsection{Observation}
\label{sec:observation}
We analyse four disc-integrated emission spectra of WASP-43b derived from a phase curve observation with \textit{JWST} MIRI in Low Resolution Spectroscopy slitless mode
collected between December 1 and 2, 2022 \citep[\textit{JWST} Transiting Exoplanet Community Early Release Science Program, JWST-ERS-1366, ][]{stevenson_transiting_2016,bean_transiting_2018,bell_nightside_2024}. 
The original observation spans 5 to 12 $\micron$ and contains a full phase curve with two eclipses and one transit, lasting 26.5 hours at a cadence of 10.34 s.
However, we only fit the observation in the 5-10.5 $\micron$ region due to unresolved systematic effects in the 10.6-11.8 $\micron$ region (see \ref{sec:shadow_effect}).
The spectra we analyse are the final fiducial spectra presented in \cite{bell_nightside_2024}, which are the mean average of the spectra reduced using four different data reduction pipelines (Eureka! v1, Eureka! v2, TEATRO, SPARTA). 
The uncertainties are set as the mean uncertainty per wavelength bin, and then adding in quadrature the root mean square between the individual reductions and the mean spectrum, as was done in \cite{bell_nightside_2024}.

\subsubsection{`Shadowed Region Effect'}
\label{sec:shadow_effect}
The original MIRI/LRS observation spanned 5 to 12 $\micron$. 
However, the observation in the 10.6-11.8 $\micron$ region was affected by the so-called `shadowed region effect' of the MIRI instrument \citep{bell_nightside_2024}, which has not yet been robustly separated from the astrophysical phase variations, making the observation in this wavelength range unreliable for this particular observation \citep[not all MIRI/LRS observations are affected, see][]{bell_nightside_2024}. 
Therefore, we exclude this wavelength region from our retrievals, in line with the retrieval analysis in \cite{bell_nightside_2024}. 
This wavelength region contains strong NH$_3$ features; thus, successfully removing the systematic effect in the future will allow us to more confidently constrain the NH$_3$ abundance of WASP-43b and other hot Jupiters.

\subsection{Parametric atmospheric model}
\label{sec:model}

\begin{table*}
\centering
\caption{Free parameters of our atmospheric model and their priors. 
Compared to \protect\cite{yang_testing_2023}, we additionally include NH$_3$, an error inflation parameter $b$, and extend our priors for the log$_{10}$ VMRs from (-8,-2) to (-10,-2) to account for the generally higher opacities in the MIRI bandpass compared to the \textit{HST}/WFC3 bandpass modelled in \protect\cite{yang_testing_2023}.
}
\label{tab:priors}
\begin{tabular}{lccc}
    \hline
    Parameter & Description & Prior  & Unit\\
    \hline
    $\delta$ & Dayside longitudinal offset  & $U$(-45,45) & degree        \\
    $\varepsilon$ & Dayside longitudinal width scaling & $U$(0.5,1.2) & dimensionless  \\
    $n$ & Dayside longitudinal variation exponent & $U$(0,2) & dimensionless \\
    $\log \kappa_{\tx{th,day}}$ & Log$_{10}$ mean infrared opacity (dayside) & $U$(-4,2) & m$^2$kg$^{-1}$  \\
    $\log\gamma_{\tx{day}}$ &  Log$_{10}$ ratio of visible and infrared opacities (dayside)  &  $U$(-4,1)      & dimensionless \\
    $\log f_{\tx{day}}$ & Log$_{10}$ heat redistribution parameter (dayside) & $U$(-4,1) & dimensionless \\
    T$_{\tx{int,day}}$ &  Internal heat flux temperature (dayside)   & $U$(100,1000)  & Kelvin              \\
    $\log \kappa_{\tx{th,nigtht}}$  & Log$_{10}$ mean infrared opacity (nightside)  & $U$(-4,2)  & m$^2$kg$^{-1}$    \\
    $\log\gamma_{\tx{night}}$ & Log$_{10}$ ratio of visible and infrared opacities (nightside) & $U$(-4,1)     & dimensionless\\
    $\log f_{\tx{night}}$ & Log$_{10}$ heat redistribution parameter (nightside)  &   $U$(-4,1)  & dimensionless  \\
    T$_{\tx{int,night}}$ &  Internal heat flux temperature (nightside)  &  $U$(100,1000) & Kelvin  \\
    $b$ & Error inflation parameter & $U$(1,5) & dimensionless   \\
    log VMR$_{\tx{H$_2$O}}$ & Log$_{10}$ volume mixing ratio of H$_2$O & $U$(-10,-2) & dimensionless   \\
    log VMR$_{\tx{CO$_2$}}$ & Log$_{10}$ volume mixing ratio of CO$_2$ & $U$(-10,-2) & dimensionless   \\
    log VMR$_{\tx{CO}}$ & Log$_{10}$ volume mixing ratio of CO & $U$(-10,-2) & dimensionless   \\
    log VMR$_{\tx{CH$_4$}}$ & Log$_{10}$ volume mixing ratio of CH$_4$ & $U$(-10,-2) & dimensionless   \\
    log VMR$_{\tx{NH$_3$}}$ & Log$_{10}$ volume mixing ratio of NH$_3$ & $U$(-10,-2) & dimensionless   \\
    \hline
\end{tabular}
\end{table*}

\begin{figure}
\center
\includegraphics[scale=0.8]{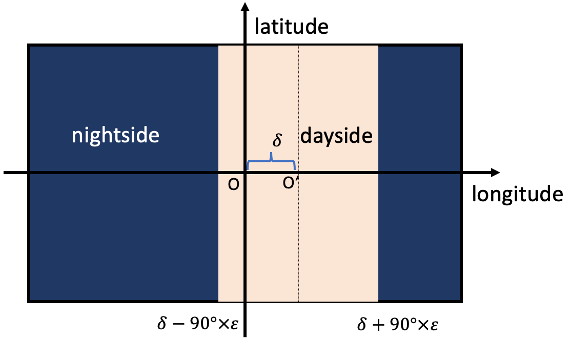}
\caption{Schematics of our parametric 2D temperature model as defined by equation (\ref{eq:Tmap}).
The model divides the atmosphere into a dayside region and a nightside region, each modelled with a representative TP profile.
The dayside central longitude and the dayside width are allowed to vary.
`O' marks the substellar point and `O$^{'}$' marks the centre of the dayside region.
While temperature is constant with longitude on isobars in the nightside region, we can parameterise the variation of temperature with longitude on the dayside.
Note that the temperature is constant with latitude and only varies with pressure and longitude (we interpret the retrieved thermal structure as a latitudinal average, see \ref{sec:model}).
}
\label{fig:model}
\end{figure}

Our atmospheric model combines a parametric 2D temperature model (model 4 from \citealt{yang_testing_2023}) with a simple chemistry model that assumes uniform gas volume mixing ratios (VMRs) in the atmospheric region probed by the observation. 
The temperature model is `2D' in the sense that temperature only varies with pressure and longitude and is held constant with latitude.
Of course, physically, we do not expect the temperature to be constant with latitude, but should instead fall off towards the poles. 
We interpret the thermal structure retrieved using this 2D model as a latitudinally-averaged thermal structure, as \cite{yang_testing_2023} found that the temperature structure retrieved from synthetic phase curve data simulated from a general circulation model (GCM) closely resembles the meridional mean thermal structure of the GCM calculated using cos(latitude) as the weight.
We choose not to model the latitudinal temperature variation since we are much more sensitive to longitudinal variation than latitudinal variation with our phase curve observation (the central longitude during the observation changes, whereas the central latitude is held constant). 
We note that the dayside latitudinal variation can be constrained with the eclipse mapping data of WASP-43b \citep{hammond_two-dimensional_2024}, discussed in \ref{sec:thermal_assumption}. 

The 2D temperature model divides the atmosphere into a dayside region and a nightside region that are each described with a representative temperature-pressure (TP) profile (see Figure \ref{fig:model}). 
The width and the central longitude of the dayside region are free parameters, modelling the effects of atmospheric circulation on heat redistribution. 
These parameters can also mimic the impact of a cloud distribution across the atmosphere, which may cause an apparent hot spot offset \citep{parmentier_cloudy_2020}.
We allow the TP profile to vary with longitude on the dayside while keeping the temperature constant with longitude on isobars on the nightside, inspired by GCM studies which show approximately constant-with-longitude meridional mean thermal structures on the nightsides of hot Jupiters \citetext{e.g., Figure 10, \citealp{kataria_atmospheric_2015};  Figure 4, \citealp{mendonca_revisiting_2018}; Figure 7, \citealp{yang_testing_2023}; Figure 2, \citealp{teinturier_radiative_2024}}.
Mathematically, the temperature $T$ in our model atmosphere at pressure $P$, longitude $\Lambda$, and latitude $\Phi$ is given by 
\begin{equation}
\label{eq:Tmap}
T(P, \Lambda, \Phi) =  
\begin{cases} 
T_{\tx{night}}(P) \text{ if }\Lambda > \delta + 90^{\circ}\times\epsilon \text{ or } \Lambda < \delta - 90^{\circ}\times\epsilon, \\
T_{\tx{night}}(P) + (T_{\tx{day}}(P)-T_{\tx{night}}(P))\cos^n(\f{\Lambda-\delta}{\varepsilon})  \\
\text{ if $ \delta + 90^{\circ}\times\epsilon \geq \Lambda \geq \delta - 90^{\circ}\times\epsilon $,}
\end{cases}
\end{equation}
where $T_{\tx{day}} (P)$ and $T_{\tx{night}}(P)$ are respectively the representative TP profiles for the dayside region and the nightside region, $\delta$ is the longitudinal deviation of the centre of the dayside region from the substellar point, and $\epsilon$ is a parameter governing the width of the dayside region. 
Our coordinate system is defined such that the sub-stellar point (where the star would be perceived to be directly overhead) is at 0 degree longitude, and the anti-stellar point is at -180 degree longitude.
The dayside region in our model is bound by the meridians $\Lambda=\delta-90^{\circ}\times\varepsilon$ and $\Lambda=\delta+90^{\circ}\times\varepsilon$, and the rest of the model is the nightside region (see Figure \ref{fig:model}). 
We reiterate that the `dayside region' in this model can deviate from the illuminated region of the planet.
The exponent $n$ of the cosine term in equation (\ref{eq:Tmap}) additionally prescribes how strongly temperatures vary with longitude on isobars on the dayside. 

Following \cite{yang_testing_2023}, we use the 1D radiative equilibrium temperature profile given by equation (29) of \cite{guillot_radiative_2010} to set $T_{\tx{day}} (P)$ and $T_{\tx{night}}(P)$. 
While the derivation of this TP profile assumes radiative equilibrium, we find in \cite{yang_testing_2023} that it is flexible enough to approximate the temperature profiles calculated by a GCM in the limited pressure range currently probed by low-resolution emission spectroscopy. 
The \cite{guillot_radiative_2010} profile is given by
\begin{equation}
\label{eq:tp_2stream}
T^4 (P)= \f{3 T_{\tx{int}}^4}{4} \Big( \f{2}{3} + \tau \Big) +  \f{3 T_{\tx{irr}}^4}{4} f \Big[ \f{2}{3} + \f{1}{\gamma\sqrt{3}} + \Big( \f{\gamma}{\sqrt{3}} - \f{1}{\gamma \sqrt{3}} \big) \tx{e}^{-\gamma \tau \sqrt{3}}\Big],
\end{equation}
where $\tau$ is the infrared optical depth given by
\begin{equation}
    \tau(P) = \f{\kappa_{\tx{th}} P}{g}. 
\end{equation}
This TP profile contains four free parameters: $\kappa_{\tx{th}}$ is the mean infrared opacity, $\gamma$ is the ratio between the mean visible and mean infrared opacities, $T_{\tx{int}}$ is the internal heat flux, and $f$ is a catch-all parameter of order unity that models the effects of albedo and the redistribution of stellar flux due to atmospheric circulation. 
We assume a negligible change in gravity $g$ in the pressure range probed by the observation for the computation of the TP profile\footnote{For our best-fit model shown in section \ref{sec:results} the change in $g$ from the highest pressure level to the lowest pressure level is about $5\%$.}, so that  $\tau$ is linear in $P$.  
$T_{\tx{irr}}$ is the irradiation temperature defined by
\begin{equation}
\label{eq:T_irr}
T_{\tx{irr}} = \Big(\f{R_{\tx{star}}}{a}\Big)^{1/2} T_{\tx{star}},
\end{equation}
where $a$ is the orbital semi-major axis, and $R_{\tx{star}}$ and $T_{\tx{star}}$ are the host star radius and temperature, respectively.
In our model, the TP profile parameters for $T_{\tx{day}} (P)$ and $T_{\tx{night}}(P)$ are set independently. 
In summary, our temperature model has eleven free parameters: eight parameters for the two representative TP profiles, plus three parameters that characterise the dayside longitudinal thermal structure (see Table \ref{tab:priors}). 

For the chemistry component of our atmospheric model, we assume the atmosphere is H$_2$/He dominated, with a fixed Jupiter-like H$_2$ to He VMR ratio at 86:14 \citep{von_zahn_helium_1996, niemann_composition_1998}.
Our model includes five spectrally active molecules whose VMRs are free parameters: H$_2$O, CO$_2$, CO, CH$_4$, and NH$_3$.
These molecules are expected at potentially detectable abundance in the atmosphere of WASP-43b from chemical modelling \citetext{\citealp{mendonca_three-dimensional_2018}; Figure 5, \citealt{venot_global_2020}; Figure 18, \citealt{baeyens_grid_2021}} and given their significant opacities in the MIRI wavelengths (Figure \ref{fig:spectral_fit}).
We assume that the VMRs of the spectrally active molecules are constant with altitude, longitude and latitude based on previous modelling works \citep[e.g.,][]{cooper_dynamics_2006, agundez_pseudo_2014, baeyens_grid_2021}, which suggest that atmospheric circulation should effectively homogenise the molecular abundances in hot Jupiters in the pressure range (10-10$^{-3}$ bar) typically probed by low-resolution emission spectroscopy. 

\begin{table}
\centering
\caption{WASP-43b stellar and planetary parameters used in this study. 
}
\label{tab:fixed_parameters}
\begin{tabular}{lcc}
    \hline
     Parameter & Value & Reference \\
    \hline
    $T_{\tx{star}}$ & 4300 K & \cite{bell_nightside_2024} \\
    $R_{\tx{star}}$ & 0.667 $R_{\sun}$  & \cite{gillon_trappist_2012} \\
    $[$Fe/H$]_{\tx{star}}$ & -0.01 dex  & \cite{gillon_trappist_2012} \\
    $a$  & 0.0153 AU & \cite{gillon_trappist_2012} \\
    $M_{\tx{plt}}$ & 2.034 $M_{\tx{Jup}}$ & \cite{gillon_trappist_2012}\\
    $R_{\tx{plt}}$ & 1.036 $R_{\tx{Jup}}$ & \cite{gillon_trappist_2012} \\
    \hline
\end{tabular}
\end{table}

The fixed stellar and planetary parameters used in this study are listed in Table \ref{tab:fixed_parameters}, and the free parameters of our model are listed in Table \ref{tab:priors}. 

\subsection{Radiative transfer}
\label{sec:radtran}

We use the correlated-k method for radiative transfer \citep{lacis_description_1991}, following \cite{irwin_nemesis_2008}.
The molecular opacity data are taken from the ExoMolOP database \citep{chubb_exomolop_2021}. 
We generate channel-integrated k-tables at the resolution of the spectra (constant 0.5 $\micron$ bins) from the original ExoMolOP k-tables, which were computed at a resolving power of R=1000.
We model the molecular absorption due to the five spectrally active molecules in our model and the collision-induced absorption due to H$_2$–H$_2$ pairs and H$_2$–He pairs.
The full list of opacity data is given in Table \ref{tab:opacities}.
We calculate the disc-integrated emission spectra at arbitrary orbital phase using the method of \cite{irwin_25d_2020} with five zenith angle quadratures, which takes into account the variation of emission angle and optical path across the atmosphere. 
The disc-integrated spectrum at a given orbital phase is computed assuming a circular, tidally locked and edge-on orbit, using the parameters in Table \ref{tab:fixed_parameters}.
The spectral calculations are implemented by the NEMESISPY package\footnote{\url{https://pypi.org/project/nemesispy}} following \cite{yang_testing_2023}, which is a Python development of the Fortran NEMESIS library \citep{irwin_nemesis_2008}.

\begin{table}
\centering
\caption{Opacity data used to calculate emission spectra in this study. 
 Apart from the He and H$_2$ collision-induced absorption opacity, all data are downloaded from the ExoMolOP database \citep{chubb_exomolop_2021}.}
\label{tab:opacities}
\begin{tabular}{lcc}
    \hline
    Molecule  &  Opacity Data  \\
    \hline
    H$_2$O  & \cite{polyansky_exomol_2018}    \\
    CO$_2$ &  \cite{yurchenko_exomol_2020}   \\
    CO      & \cite{li_rovibrational_2015}    \\
    CH$_4$  & \cite{yurchenko_hybrid_2017}  \\
    NH$_3$  & \cite{coles_exomol_2019}  \\
    He                 &   \cite{borysow_collision-induced_1989} \&       \cite{borysow_collision-induced_1989-1}  \\
    H$_2$              &    \cite{borysow_collision-induced_1989} \&          \cite{borysow_collision-induced_1989-1}  \\
    \hline
\end{tabular}
\end{table}

\subsection{Retrieval set-up}
\label{sec:retrieval}
We perform retrievals on the emission spectra of WASP-43b at four different orbital phases simultaneously. 
Following \cite{bell_nightside_2024}, the spectra are binned to eleven 0.5$\micron$ wavelengths bins in the 5-10.5$\micron$ wavelength range, at orbital phase 0 (nightside), 0.25 (observer facing the evening terminator), 0.5 (dayside) and 0.75 (observer facing the morning terminator). 
We fit the four spectra simultaneously using spectra calculated from our atmospheric model described in \ref{sec:model} via the radiative transfer procedure described in \ref{sec:radtran}. 
We convert planetary emission to planet-to-star flux ratio using a stellar spectrum computed using a PHOENIX stellar model \citep{allard_model_1995, hauschildt_nextgen_1999, husser_new_2013}, generated assuming an effective temperature of 4300 K, a surface gravity of log(g)= 4.50, and a solar metallicity.

The atmospheric model is defined from 20 to $10^{-5}$ bar, on 40 points equally spaced in log pressure. 
We retrieve the eleven free temperature model parameters, the five molecular VMRs, and an error inflation parameter described in \ref{sec:error_inflation}. 
The priors of these parameters are listed in Table \ref{tab:priors}. 
We find the posterior distribution of these parameters given the data using the Nested Sampling algorithm \citep{feroz_multimodal_2008} as implemented by the PyMultiNest software package \citep{buchner_x-ray_2014}, using 1000 sampling live points. 
For each molecule, we assess the significance of its detection via Bayesian model comparison by performing five additional retrievals, each omitting one of the five molecules in our fiducial model. 
The detection significance of a molecule can be calculated by computing the Bayes factor of the fiducial model over the model without the particular molecule, following \cite{trotta_bayes_2008} and \cite{benneke_how_2013}.

\subsubsection{Error inflation}
\label{sec:error_inflation}
We include an error inflation parameter to account for underestimated measurement and forward modelling uncertainties, following the approach used in, for example, \cite{line_uniform_2015} and \cite{bell_nightside_2024}.
We also refer the reader to \cite{irwin_nemesis_2008}, where the forward modelling error is treated in the optimal estimation framework.
The error inflation is implemented by modifying the likelihood term in Bayesian inference with a constant-with-wavelength multiplicative factor. 
The log likelihood function with error inflation is given by
\begin{equation}
\ln L(\pmb{D}|\pmb{\theta}) =  -\f{1}{2} \sum^{n}_{i=1} \f{\left(D_i - F_i(\pmb{\theta}) \right)^2}{s_i^2} - \f{1}{2} \ln (2\pi s_i^2),
\label{eq:loglike}
\end{equation}
where $\pmb{D}$ is the set of point data points in the observed spectra, $F_i(\pmb{\theta})$ is the $i$th data point in the model spectra given model parameters $\pmb{\theta}$, and $s_i$ is defined as 
\begin{equation}
s_i = b * \sigma_i,
\label{eq:error}
\end{equation}
where $\sigma_i$ is the uncertainty for the $i$th data point, and $b$ is a free error inflation parameter between 1 and 5.

\section{Results}
\label{sec:results}

\begin{figure*}
\includegraphics[scale=0.68]{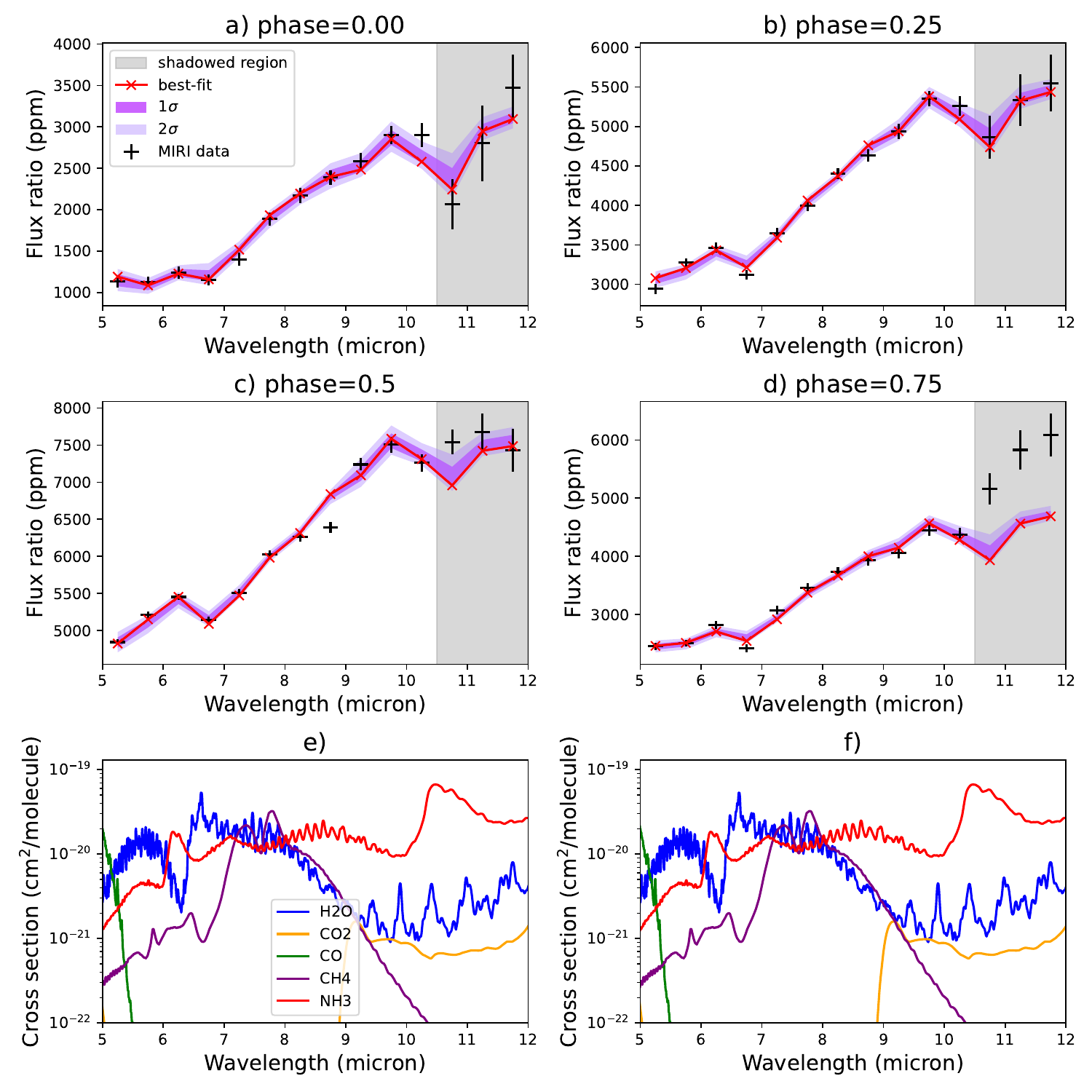}
\caption{a) - d): Best-fit model spectra at the four orbital phases calculated from the maximum a posteriori parameters. 
We plot the 1$\sigma$ and 2$\sigma$ central credible intervals with dark and light purple shading, respectively. 
The data in the grey shaded region (10.5-12 $\micron$) are affected by the `shadowed region' systematics and are not included in our retrievals (see \ref{sec:shadow_effect}).  
Interestingly, our best-fit model, when extended to 12 $\micron$, is consistent with the current reduction of the `shadowed region' data for all phases except at phase $0.75$ (see \ref{sec:spectral_fit} for why we think phase $0.75$ is worst affected by the `shadowed region effect').
Since NH$_3$ has strong spectral features in the `shadowed region', a reliable reduction of the `shadowed region effect' in the future can refine the abundance constraints on NH$_3$.
Note the absorption feature at the 8.75 $\micron$ bin at phase $0.5$, which our model cannot explain.
e) - f): Cross sections of the spectrally active molecules included in our retrievals, computed at a spectral resolution of R$=1000$ at 1 bar pressure and 1500 K temperature. 
f) is identical to e) except for the omission of legend for ease of reference.
The spectral features between $\sim$5 and $\sim$8 $\micron$ are indicative of H$_2$O molecules, whereas the absence of spectral features between $\sim$7 and $\sim$9 $\micron$ allows us to rule out CH$_4$ at high abundance.
While NH$_3$ has strong opacities throughout the MIRI/LRS bandpass, its retrieved abundance is about two orders of magnitude lower than H$_2$O. 
This means that the spectral features of H$_2$O would dominate over those of NH$_3$ in the wavelength regions where H$_2$O has higher or similar opacity compared to NH$_3$.
The notable exception is the wavelength region greater than $\sim$8 $\micron$, where the H$_2$O opacity is generally much lower than the NH$_3$ opacity. 
The constraints on CO are driven by the 5.25 $\micron$  bin; the absence of CO would increase the planetary flux at the 5.25 $\micron$  bin relative to the 5.75 $\micron$ bin. 
The constraints on CO$_2$ are driven by the data in 9-10.5 $\micron$ region.}
\label{fig:spectral_fit}
\end{figure*}

\begin{figure*}
\includegraphics[scale=0.5]{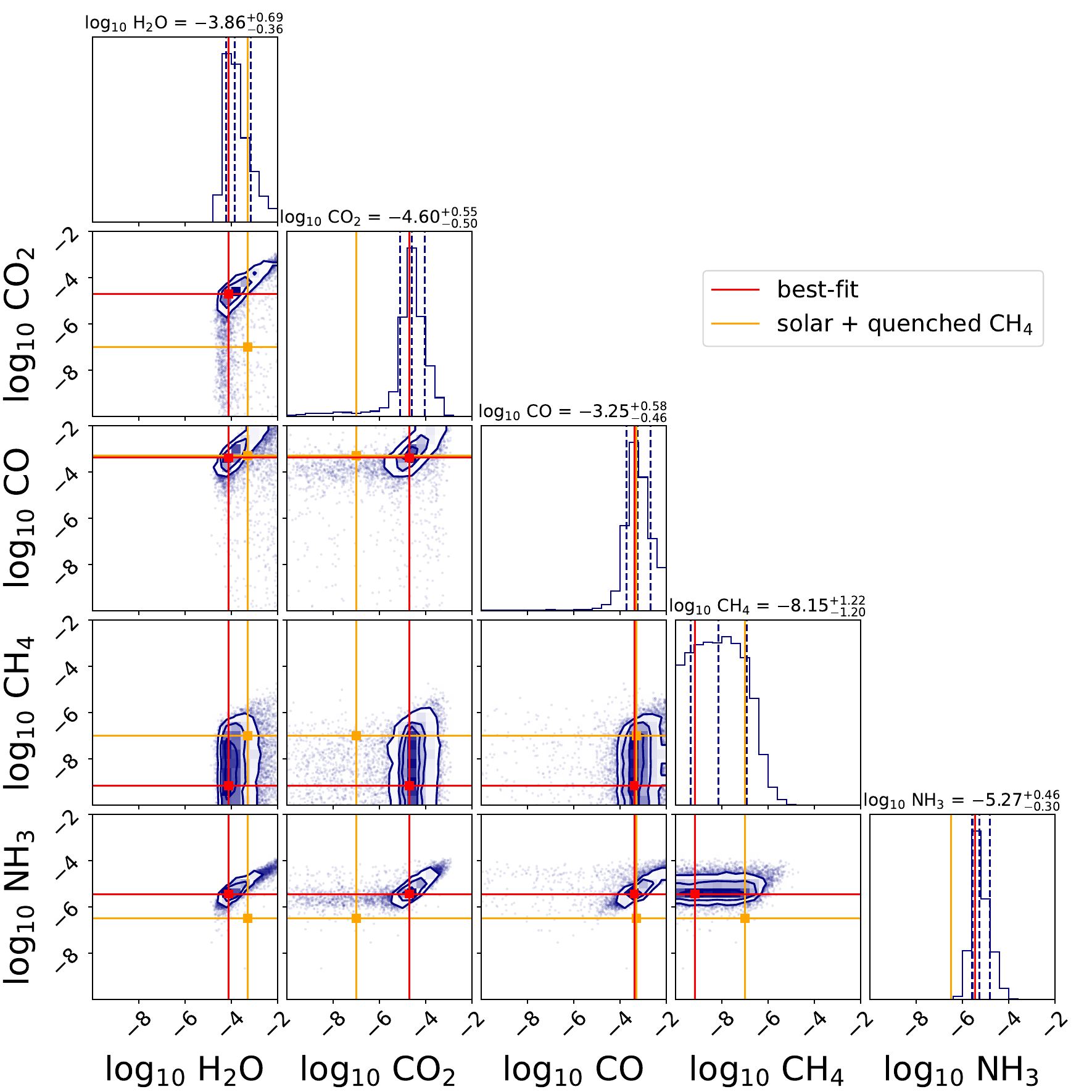}
\caption{Posterior distribution of the molecular abundances.
The histograms on the diagonal show the marginalised posterior distributions of each molecular VMR, where the blue dashed lines give the 16\%, 50\% and 84\% percentiles (the percentile values are given at the top of each panel).  
The off-diagonal plots show the joint posterior distributions of any pairs of parameters.
The red solid lines mark our retrieved best-fit parameters (maximum a posteriori parameters).
The orange solid lines mark the expected log$_{10}$ VMR of the molecules in a solar metallicity atmosphere, taking horizontal and vertical mixing into account: -3.3 for H$_2$O and CO, -7 for CO$_2$ and CH$_4$, and -6.5 for NH$_3$ \protect\citep{venot_global_2020, baeyens_grid_2021}.
While the abundances of H$_2$O, CO and CO$_2$ in the chemical models of \protect\cite{baeyens_grid_2021} and \protect\cite{venot_global_2020} are approximately constant with altitude, there are some vertical variations in the abundances of CH$_4$ and NH$_3$ in the pressure region probed by our observation. 
We take the `solar abundances' of CH$_4$ and NH$_3$ to be roughly the values at 1 bar.
}
\label{fig:vmr_posterior}
\end{figure*}

We describe the results of our retrievals in this section. 
In \ref{sec:spectral_fit}, we show that our parametric atmospheric model can explain the observation. 
In \ref{sec:abundance}, we show that there is statistically significant evidence of H$_2$O, NH$_3$, CO, weak evidence of CO$_2$, and no detectable level of CH$_4$. 
In \ref{sec:thermal_structure}, we plot the transmission weighting function of our best-fit model atmosphere and the best-fit thermal structure.
We compare our results to past \textit{HST} + \textit{Spitzer} observations in \ref{sec:HST} and briefly discuss the implication of our results on the formation history of WASP-43b in \ref{sec:formation}.
We investigate the implications of our retrieved abundances and thermal structure on the atmospheric chemistry in \ref{sec:1D_chem}. 
The posterior distribution of all our model parameters is shown in the appendix (\ref{fig:full_posterior}), alongside a comparison of our retrieved H$_2$O abundance to a list of previous studies (\ref{fig:compare_all}).

\subsection{Spectral fit}
\label{sec:spectral_fit} 
We plot the best-fit model spectra against the observed emission spectra in Figure \ref{fig:spectral_fit}.
The best-fit model spectra are calculated from the maximum a posteriori parameters (which in our case are the same as the maximum likelihood parameters), and we propagate the $1\sigma$/$2\sigma$ central credible intervals to show the spread of our posterior distribution. 
Our best-fit model can fit almost all observed data points within the original uninflated error bars in \cite{bell_nightside_2024}, which is also reflected by the fact that the error inflation parameter we retrieve is only around 1.65 (see \ref{sec:error_inflation} and \ref{fig:full_posterior}). 
One notable exception is the 8.75 $\micron$ bin at phase 0.5, which cannot be accounted for by any of the opacity sources in our atmospheric model (see \ref{sec:chemistry} for more discussion).
For ease of reference, we include plots of the cross sections of the molecules included in our retrievals in the MIRI/LRS wavelength range in Figure \ref{fig:spectral_fit}. 
Note that the absorption due to each molecule is also weighted by its VMR.
Chemical models predict that the VMRs of H$_2$O and CO in the atmosphere of WASP-43b are at least two orders of magnitude higher than the VMRs of CO$_2$, CH$_4$ and NH$_3$ \citep[e.g.,][]{venot_global_2020}, so we expect H$_2$O opacity to dominate in the 5-8 $\micron$ region.
On the other hand, NH$_3$ opacity should be comparable to the H$_2$O opacity in the 8-10.5 $\micron$ region, allowing us to constrain the abundance of NH$_3$.
For the carbon-bearing molecules, as we can see from panel e) of Figure \ref{fig:spectral_fit}, the constraints on CO could only come from  the 5.25 $\micron$ bin, whereas the constraints on CO$_2$ would be driven by the 9-10.5 $\micron$ region.
We note that a more robust determination of the relative contribution of different wavelength bins to the molecular detections can be achieved by a leave-one-out cross-validation analysis, as detailed in \cite{welbanks_application_2023}.

We reiterate that our retrievals exclude the data in the `shadowed region', which is shaded in grey in Figure \ref{fig:spectral_fit}.
Therefore, we do not fit for the wavelength region longer than 10.5 $\micron$, so our retrieval analysis excludes most of the strong NH$_3$ features centred around 10.5 $\micron$.
Interestingly, our best-fit model, when extended to the 10.5-12 $\micron$ `shadowed region', seems to be consistent with the current reduction of the `shadowed region' data for three out of four orbital phases (phases 0, 0.25, and 0.5).
Because the severity of the `shadowed region effect' (see \ref{sec:shadow_effect}) decreased as a function of time since the beginning of the MIRI observation, we expect phases 0.5 and 0.75 to be more severely affected as the observation started just before a secondary eclipse (where phase=0.5).
However, since phase 0.5 was observed again at the end of the observation, we may expect it to be less affected than phase 0.75, which was only observed once.
On the other hand, we would expect phases 0 and 0.25 to be least affected since they were observed later when much of the `shadowed region effect' had decayed away. 
Therefore, if we assume that our best-fit model is representative of the true atmosphere of WASP-43b and the `shadowed region' data is more reliable at phases 0, 0.25, and 0.5 than at phase 0.75, then we expect our best-fit model to be more consistent with the `shadowed region' data at phases 0, 0.25 and 0.5 than at phase 0.75.
As shown in Figure \ref{fig:spectral_fit}, our best-fit model agrees well with the data in the `shadowed region' for phase 0 and phase 0.25, and slightly less so at phase 0.5.
In contrast, our best-fit model does not agree with the `shadowed region' data at phase 0.75.
Considering the time dependence of the `shadowed region effect', the data in the `shadowed region' is consistent with our best-fit model and the presence of NH$_3$. 
The important caveat is that the poorly fit systematic noise could unexpectedly impact the data reduction in \cite{bell_nightside_2024}, so we do not want to draw any definitive conclusions from the affected data until the systematics issue is fully resolved. 
Further observations of WASP-43b using other \textit{JWST} instruments at shorter wavelengths (such as the scheduled NIRSpec observation, GTO 1224, PI: S. Birkmann), re-observations of WASP-43b using MIRI after the (still unknown) cause of the `shadowed region effect' has been mitigated, or a robust reduction of the `shadowed region effect' can all be used to confirm our NH$_3$ signal and tighten the constraints on its abundance.   

\subsection{Abundance constraints}
\label{sec:abundance}

\begingroup
\setlength{\tabcolsep}{10pt} 
\renewcommand{\arraystretch}{1.3} 
\begin{table}
\centering
\caption{Detection significance and abundance constraints of molecules included in our retrieval model.
For all molecules apart from CH$_4$, we quote the medians of the marginalised posterior distributions,  with the uncertainties showing the 1$\sigma$ central credible intervals.
For CH$_4$, we quote the 95\% upper limit.}
\label{tab:results_sigma}
\begin{tabular}{lccc}
    \hline
     Molecule & Detection Significance &  log$_{10}$(VMR)       \\
    \hline
     H$_2$O &  6.5$\sigma$ & $-3.86^{+0.69}_{-0.36}$  \\
    CO$_2$  & 2.5$\sigma$  & $-4.60^{+0.55}_{-0.50}$ \\
    CO      & 3.1$\sigma$  & $-3.25^{+0.58}_{-0.46}$ \\
    CH$_4$ &  NA           & <-6.39 \\
    NH$_3$ & 4$\sigma$  & $-5.27^{+0.46}_{-0.30}$ \\
    \hline
\end{tabular}
\end{table}
\endgroup

We show the retrieved posterior distribution of the molecular VMRs in Figure \ref{fig:vmr_posterior}.
We mark our best-fit model with red solid lines and a solar-metallicity GCM-based chemistry model in orange solid lines for reference (R. Baeyens, private communication). 
Using the Bayesian evidence ratio test described in \ref{sec:retrieval}, we find statistically significant evidence of H$_2$O at 6.5$\sigma$, NH$_3$ at 4$\sigma$, and CO at 3.1$\sigma$.
We find weak evidence of CO$_2$ (2.5$\sigma$) and no detectable level of CH$_4$ (95\% VMR upper limit at $10^{-6.39}$). 
We note that our analysis is one of the first studies to report evidence of NH$_3$ in the atmosphere of an exoplanet \citep{macdonald_hd_2017, macdonald_signatures_2017, giacobbe_five_2021, dyrek_so2_2024}, and to our knowledge, the first to report evidence of NH$_3$ by analysing emission spectra.
We list the detection significance and the abundance constraints (given as 1$\sigma$ central credible intervals) in Table \ref{tab:results_sigma}.

\begin{figure}
\center
\includegraphics[scale=0.3]{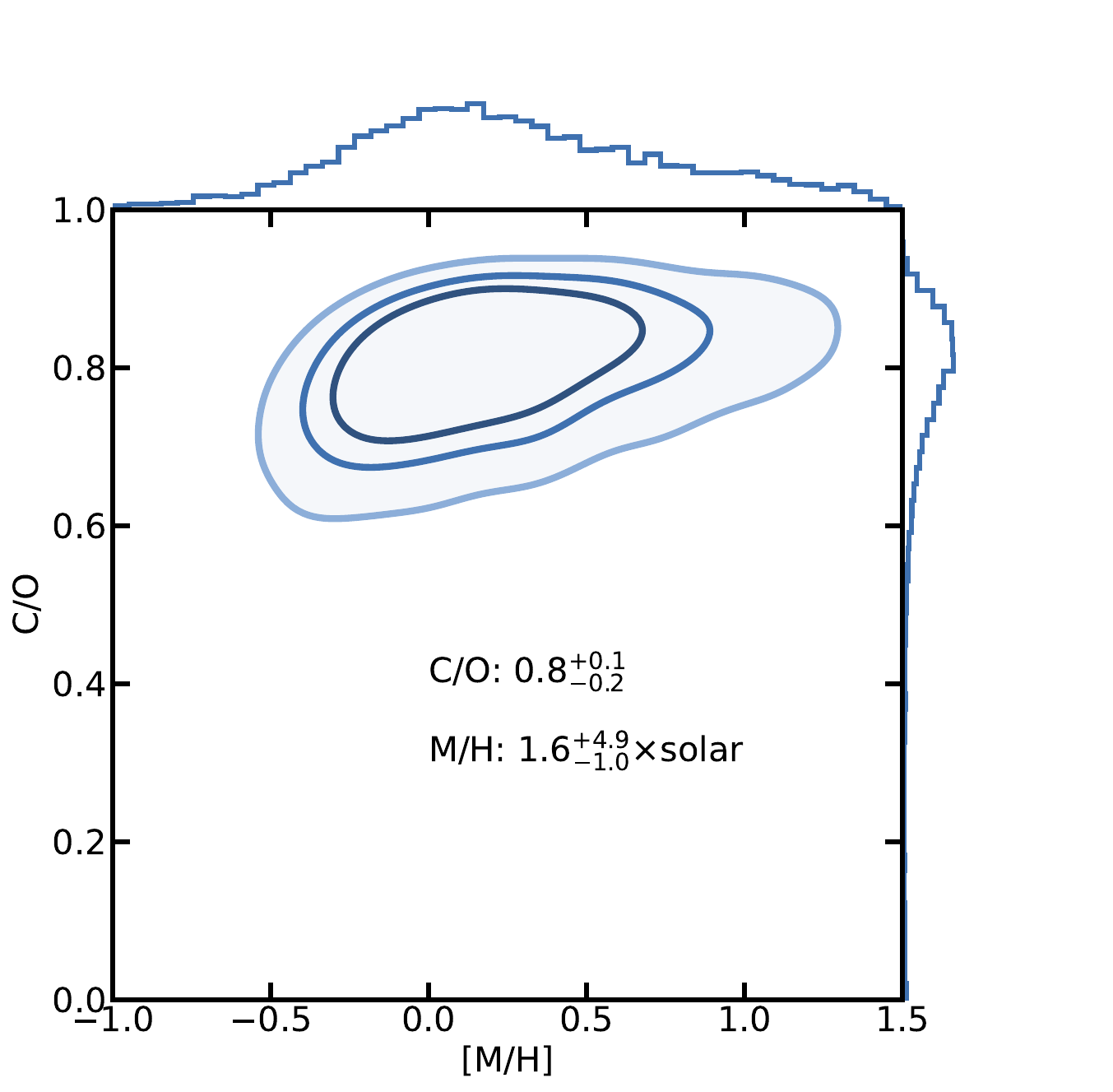}
\caption{Joint distribution of metallicity and C/O ratio derived from the posterior distribution of all molecular VMRs in our model.
}
\label{fig:Met}
\end{figure}

As shown in Figure \ref{fig:vmr_posterior},  the retrieved H$_2$O and CO abundances are broadly consistent with a solar metallicity atmosphere. 
We constrain the metallicity and C/O ratio of WASP-43b using the posterior distribution of the molecular VMRs in Figure \ref{fig:Met}, taking the reference solar abundances from \cite{lodders_solar_2010-2}.
Note that our metallicity estimate is calculated taking into account all of the molecules included in our retrievals, as listed in Table \ref{tab:results_sigma}, thus taking into account O, C, and N elemental abundances.
We tentatively estimate the metallicity of WASP-43b at $1.6^{+4.9}_{-1.0}\times$ solar and its C/O ratio at $0.8^{+0.1}_{-0.2}$.
We find that our retrieved abundance constraints, and the inferred metallicity and C/O ratio are very similar to the results of \cite{lesjak_retrieval_2023}, which were based on high-solution dayside emission spectra observed with VLT/CRIRES+, thus providing validation to our results with an independent observational method.
We discuss the implications of our retrieved abundances on the formation of WASP-43b in section \ref{sec:formation}.

We caution the reader that our metallicity and C/O ratio estimations are only tentative for the following reasons. 
Firstly, our CO constraints come only from the shortest 5.25 $\micron$ wavelength bin (Figure \ref{fig:spectral_fit}).
Since the shortest 1-2 wavelength bins are much more strongly illuminated, they may suffer more from poorly modelled non-linearity effects \citep{argyriou_brighter-fatter_2023} in the data reduction stage of \cite{bell_nightside_2024}.
Furthermore, the detection significance of the carbon-bearing molecules are relatively low, as the MIRI/LRS wavelengths are not optimal for constraining the abundances of CO and CO$_2$.
The scheduled NIRSpec observation of WASP-43b (GTO 1224, PI: S. Birkmann) covering the 2.9-5.2 $\micron$ region will more robustly constrain the abundances of both CO and CO$_2$, which will lead to a more accurate estimation of C/O ratio.
We further note that cloud formation can also bias our estimated metallicity and C/O ratio.
As pointed out by \cite{lodders_atmospheric_2002} and \cite{woitke_equilibrium_2018}, the formation of silicate clouds can take oxygen away from gas-phase molecules. 
Therefore, if there is a significant level of silicate cloud formation on the nightside of WASP-43b,
then our C/O ratio estimation would be an overestimate and our metallicity estimation would be an underestimate.

\begin{figure}
\centering
\includegraphics[scale=0.5]{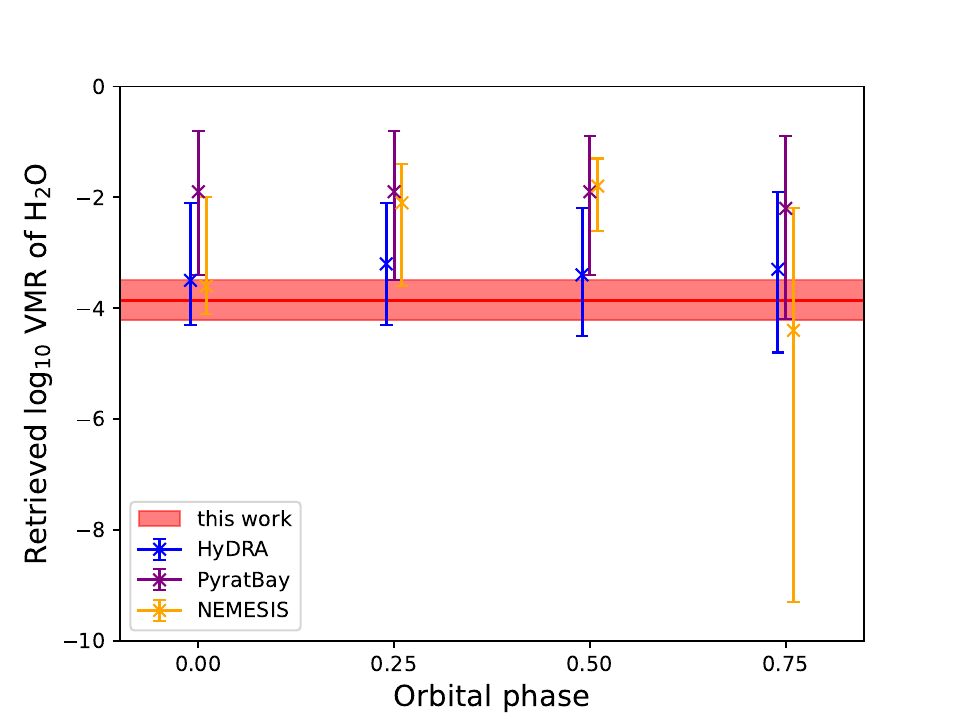}
\caption{Comparison of the retrieved constraints on H$_2$O VMR in this work to those retrieved in \protect\cite{bell_nightside_2024}.
The horizontal red solid line shows the median of the marginalised posterior distribution, and the red shaded region around it shows the 1$\sigma$ central credible interval.
We only include the retrievals in which the abundance of H$_2$O was a free parameter. }
\label{fig:compare_bell24}
\end{figure}

We now compare our abundance constraints to those in \cite{bell_nightside_2024}, which were obtained by performing retrievals on the spectrum at each orbital phase separately and using a dilution parameter to account for spatial inhomogeneity in thermal structure \citep{taylor_understanding_2020}.
We confirm the detection of H$_2$O at a much higher significance level (6.5$\sigma$) than the analysis in \cite{bell_nightside_2024}, which reported significance ranging from 1.8$\sigma$ to 4.1$\sigma$ depending on the orbital phase and the retrieval pipeline. 
Our constraints are much more precise than those in \cite{bell_nightside_2024} and are most consistent with their HyDRA retrieval pipeline results.
For the retrievals in \cite{bell_nightside_2024} that included H$_2$O abundance as a free parameter, we note that all three retrieval pipelines show results consistent with our model assumption that the H$_2$O abundance is constant across all orbital phases.
We also confirm the non-detection of CH$_4$ in \cite{bell_nightside_2024} at a 95\% upper limit of $10^{-6.39}$.
In addition to tightening the abundance constraints reported in \cite{bell_nightside_2024}, we are able to report statistically significant evidence of two additional molecules, namely NH$_3$ and CO.
Our results demonstrate the power of phase-resolved spectroscopy in characterising the chemistry of hot Jupiters in detail, as well as the importance of a global model in interpreting phase curves.

\subsection{Thermal structure}
\label{sec:thermal_structure}

\begin{figure}
\center
\includegraphics[scale=0.3]{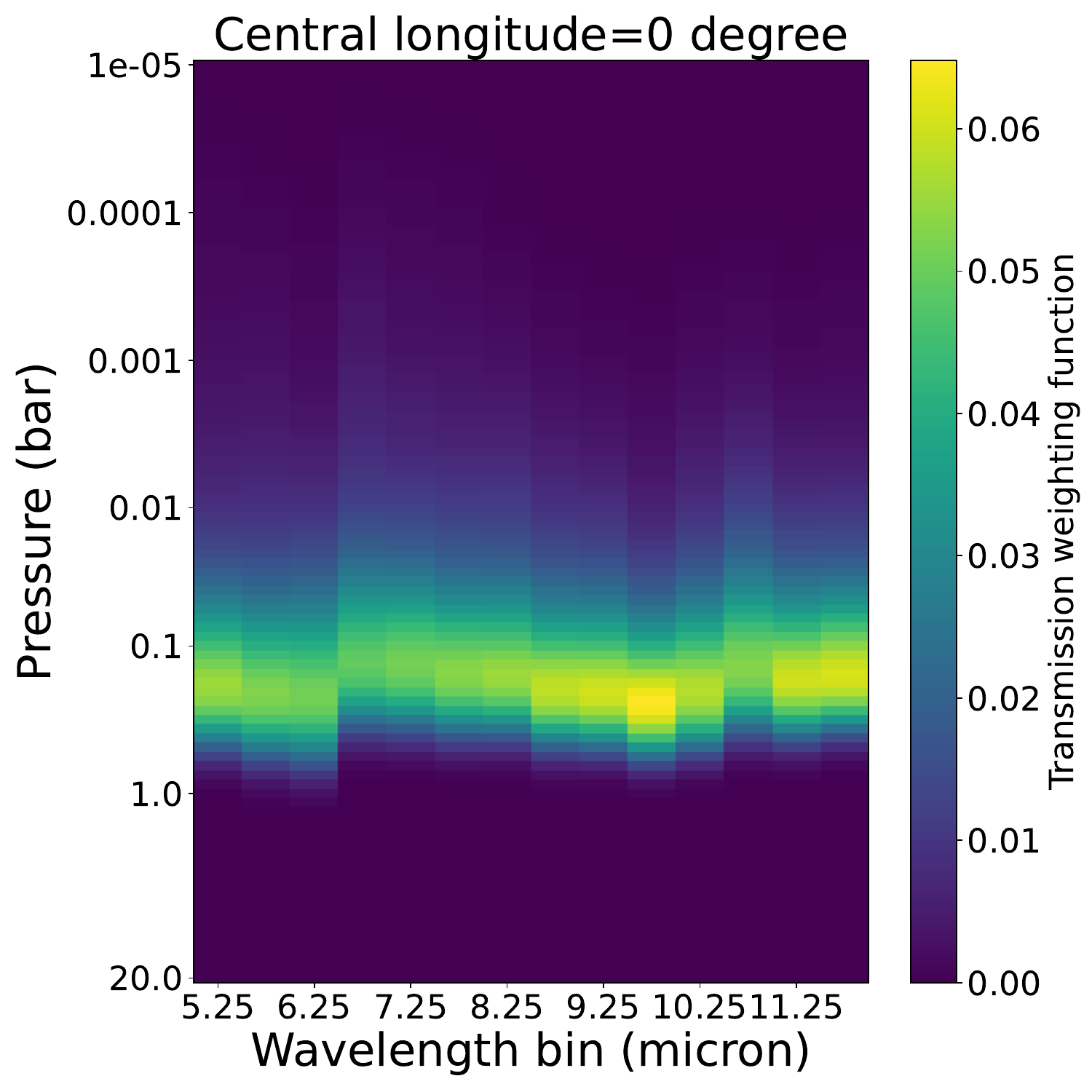}
\caption{Transmission weighting function of the best-fit retrieved model at the substellar point.
}
\label{fig:twf_day}
\end{figure}

\begin{figure}
\center
\includegraphics[scale=0.3]{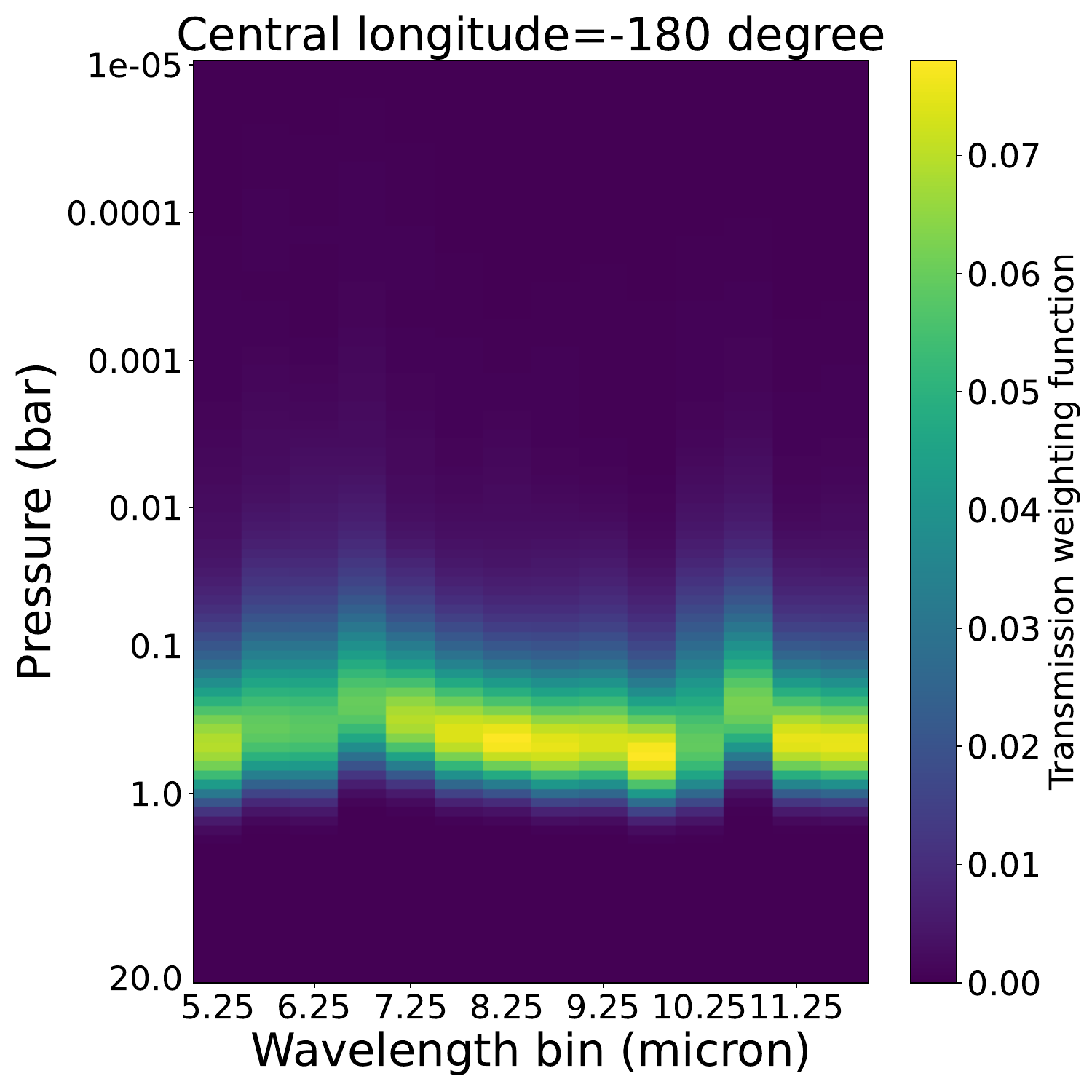}
\caption{Transmission weighting function of the best-fit retrieved model on the nightside.
}
\label{fig:twf_night}
\end{figure}

\begin{figure}
\center
\includegraphics[scale=0.5]{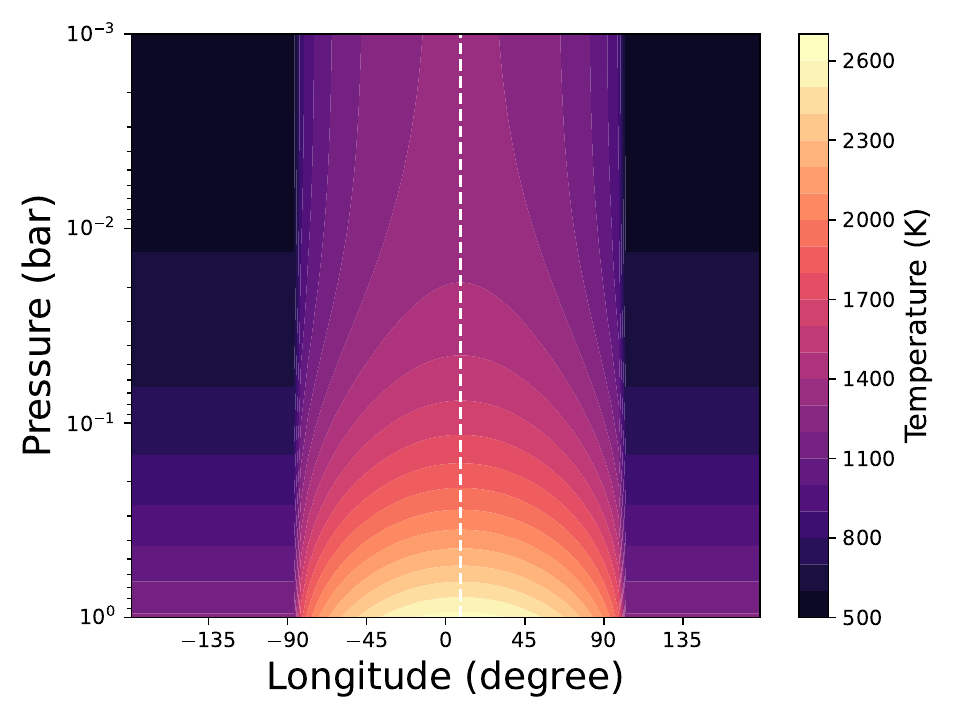}
\caption{Best-fit 2D thermal structure calculated from the maximum a posteriori parameters. 
The white vertical dashed line marks the position of the hot spot offset.
}
\label{fig:tp_2D}
\end{figure}

We plot the transmission weighting function of our best-fit model at the substellar point and the nightside in Figure \ref{fig:twf_day} and Figure \ref{fig:twf_night}, respectively. 
We see that our retrieval is primarily sensitive to pressures inside the 1 bar to 1 millibar pressure range.
We plot the retrieved best-fit temperature profile as a function of pressure and longitude in the 1 bar to 1 millibar pressure range in Figure \ref{fig:tp_2D}, as the profile outside this pressure range is not directly constrained by the data and is rather set by our parameterisation.
The retrieved temperature profile should be interpreted as a latitudinally-averaged thermal structure: \cite{yang_testing_2023} find that the retrieved thermal structure from GCM-simulated phase curves closely resemble the latitudinally-averaged GCM thermal structure using cos(latitude) as the weight.\footnote{
In \cite{irwin_25d_2020}, the authors also found that the retrieved thermal structure from GCM-simulated phase curves using a different method closely resembles the latitudinally-averaged GCM thermal structure using cos(latitude) as the weight.}
This weighting scheme is also commonly used in GCM studies to calculate meridionally averaged temperature structure \citep[e.g.,][]{ kataria_atmospheric_2015, mendonca_revisiting_2018, carone_equatorial_2020}.

The best-fit thermal structure, as shown in Figure \ref{fig:tp_2D}, consists of a hot dayside region and a cooler nightside region, where the centre of the hot dayside region is moderately offset to the east.  
The eastward shift of our retrieved thermal structure is due to the fact that the peak of the phase curve occurred before the secondary eclipse \citep{bell_nightside_2024}, and similar phase curve maximum offsets have been observed in numerous hot Jupiters \citep[e.g.,][Figure 4]{Parmentier2017}. 
The eastward shift has been predicted and explained by GCM studies \citep{showman_atmospheric_2002, showman_atmospheric_2009, showman_equatorial_2011}, which proposed that strong equatorial super-rotating (eastward) jets would develop in the atmospheres of hot Jupiters and act to transport heat eastwards, so the hottest region of the atmosphere is displaced eastward of the substellar point. 
This offset is also seen in numerous GCM simulations specific to the planetary parameters of WASP-43b \citep{kataria_atmospheric_2015, mendonca_revisiting_2018, mendonca_three-dimensional_2018, carone_equatorial_2020, schneider_exploring_2022, teinturier_radiative_2024}.
We retrieve an eastward hotspot offset of $8.68^{+0.96}_{-1.14}$ degrees, consistent with the $7.34^{+0.38}_{-0.38}$ degrees eastward offset in \cite{bell_nightside_2024}\footnote{We note that the offset of the phase curve maximum does not necessarily correspond to the longitudinal offset of the hottest atmospheric region if there are clouds, latitudinal thermal variation or more complex longitudinal thermal variation than permitted by our model \citep{parmentier_cloudy_2020}.}.
We note that our best-fit model in Figure \ref{fig:tp_2D} resembles some GCM simulations of WASP-43b, for example, Figure 10 in \cite{kataria_atmospheric_2015} and Figure 4 in \cite{mendonca_revisiting_2018}.

\begin{figure}
\center
\includegraphics[scale=0.6]{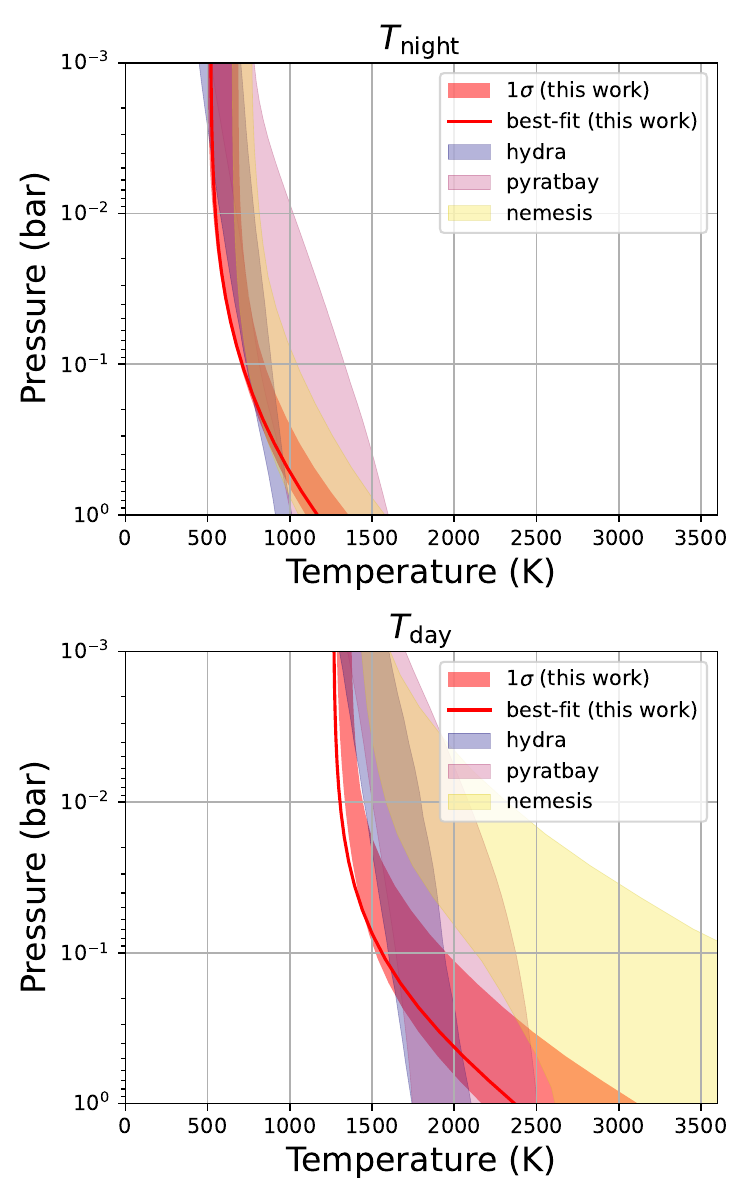}
\caption{
Comparison of the retrieved representative dayside/nightside TP profiles in this work to the retrieved dayside/nightside TP profiles in \protect\cite{bell_nightside_2024}.
The best-fit model (red line) is calculated from our maximum a posteriori parameters, whereas the 1$\sigma$ credible intervals (red shaded areas) are propagated from the 1$\sigma$ central credible intervals of our posterior distribution.
The blue, purple, yellow shaded regions represent the 1$\sigma$ credible intervals of the TP profiles retrieved using three different retrieval pipelines in Fig. 4 of \protect\cite{bell_nightside_2024}. 
We only include the retrievals in which the chemical abundances were free parameters. 
}
\label{fig:tp_profiles}
\end{figure}

We plot the retrieved representative nightside and dayside TP profiles $T_{\text{night}}$ and $T_{\text{day}}$ in Figure \ref{fig:tp_profiles}.
We find large day-night temperature contrasts of at least 700 K at all pressure levels. 
As shown by Figure \ref{fig:tp_profiles}, our retrieved temperature profiles are broadly consistent with the TP profiles retrieved in \cite{bell_nightside_2024} (see their Figure 4).
While we do not model clouds in our retrievals, past studies \citep[e.g.,][]{burningham_retrieval_2017, molliere_retrieving_2020} have shown that flexible TP profiles can mimic the spectral effects of clouds by becoming more isothermal. 
Our retrieved large day-night temperature contrast and relatively isothermal nightside TP profile are consistent with the presence of clouds on the nightside, which echo the results of \cite{bell_nightside_2024} that cloudy GCMs can match the nightside emission of WASP-43b better than cloudless GCMs.
However, we cannot rule out or confirm the presence of nightside clouds directly with our model.

\subsection{Comparison with \textit{HST}/WFC3 and \textit{Spitzer} observations}
\label{sec:HST}

\begin{figure}
\includegraphics[scale=0.34]{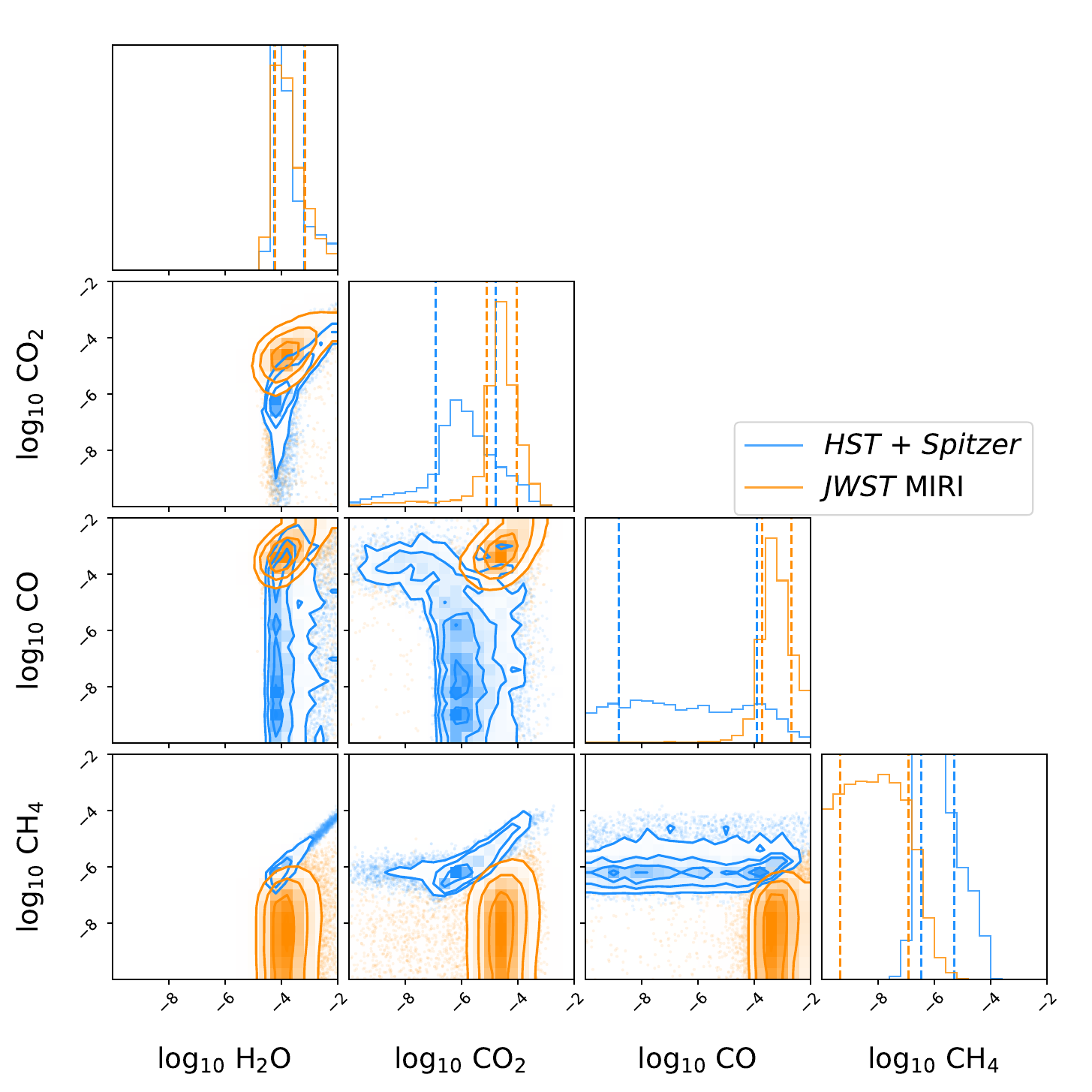}
\caption{Comparison of posterior distributions from (1) \textit{HST}/WFC3 and \textit{Spitzer}/IRAC data set analysed in \protect\cite{yang_testing_2023}; (2) \textit{JWST} MIRI/LRS data set analysed in this work. 
}
\label{fig:compare_hst}
\end{figure}

We compare the constraints on chemical abundances retrieved in this work to the abundance constraints retrieved from the \textit{HST}/WFC3 + \textit{Spitzer}/IRAC data set \citep{stevenson_spitzer_2017} in a previous study \citep{yang_testing_2023} in Figure \ref{fig:compare_hst}.
The abundance constraints on H$_2$O retrieved from the MIRI/LRS data set are entirely consistent with those retrieved from the \textit{HST}/WFC3 + \textit{Spitzer}/IRAC data set. 
This agreement boosts our confidence in the H$_2$O detection in the atmosphere of WASP-43b.
Moreover, as shown in \cite{yang_testing_2023} and \cite{irwin_25d_2020}, the \textit{HST}/WFC3 observation probes deeper in the atmosphere (to pressure levels as high as $\sim$10 bar) than the MIRI/LRS observation analysed in this work.  
The consistent H$_2$O abundance retrieved from the two data sets reaffirms the assumption that the abundance of H$_2$O is constant with pressure in the observable pressure levels of hot Jupiters.
We compare our retrieved H$_2$O abundance to more previous studies in appendix \ref{appendix:comparison}.

\subsection{Implication on formation}
\label{sec:formation}
The composition of the protoplanetary discs in which planets form is a function of time and location, so the final composition of the planets may be used to constrain their formation and migration history.
As discussed in \ref{sec:abundance}, our results tentatively suggest that the atmosphere of WASP-43b has a metallicity of 0.6-6.5$\times$ solar, with a super-solar C/O ratio of $0.8^{+0.1}_{-0.2}$. 
We now discuss the implications of these estimates on the formation history of WASP-43b.
The retrieved metallicity is consistent with the mass-metallicity relation in which more massive planets are expected to have lower metallicities (e.g., Figure 4, \citealt{kreidberg_precise_2014}; Figure 2, \citealt{thorngren_connecting_2019}; Figure 12, \citealt{cridland_connecting_2019}; Figure 3, \citealt{welbanks_massmetallicity_2019}), as they can accrete and retain more gas directly from the disc and thus lower their metallicity.
The retrieved C/O can inform us about the relative importance of solid accretion to gas accretion during formation, as the solid icy materials in protoplanetary discs tend to be O-rich, while the gas tends to be more C-rich. 
As mentioned above, since more massive planets such as WASP-43b ($\sim$2 Jupiter mass) accrete more gas from the disc, their atmospheres should naturally have super-solar C/O ratios. 
If we combine our retrieved metallicity and C/O ratio, we can broadly constrain the formation environment of WASP-43b. 
As pointed out by \cite{cridland_connecting_2019-1}, chemical processing of carbon-rich dust grains within the protoplanetary discs \citep{anderson_destruction_2017, gail_spatial_2017, klarmann_radial_2018} is a significant source of gaseous carbon and can impact the C/O ratios of the giant planets that form within.
Our retrieved metallicity and C/O ratio favours the `ongoing' model in  \cite{cridland_connecting_2019-1}, where ongoing chemical processes continually erode carbon from dust grains in the protoplanetary disc (Figure 6, \citealt{cridland_connecting_2020}).
Additionally, according to \cite{madhusudhan_atmospheric_2017}, who studied the link between pebble accretion and C/O ratio, our retrieved C/O ratio is more consistent with a planet that formed outside the water ice line and then migrated through the disc, but probably not outside the CO$_2$ iceline as it would lead to substantially subsolar metallicity.
The hypothesis that WASP-43b formed beyond the water iceline is supported by \cite{mordasini_imprint_2016}, who argued that planets formed within the water iceline should have C/O ratio less than 0.2 assuming a standard disc chemistry model.
The analysis presented here is a simplistic one since the impact of non-standard disc chemistry on the atmospheric composition of hot Jupiters can be highly uncertain, and we cannot rule out the possibility that the C/O ratio we retrieved is not representative of the entire atmosphere (see the discussion in \ref{sec:abundance})
. 
In addition to the C/O ratio, the N/O ratio has also gained interest as a way of connecting giant planet observations to their formation scenarios \citep{piso_role_2016, cridland_connecting_2020,  turrini_tracing_2021,ohno_jupiters_2021,notsu_molecular_2022},though additional modelling beyond the scope of this work is required to estimate the N/O ratio from our retrieved NH$_3$ abundance, as the abundance of N$_2$ may be significant but is not constrained by our model \citep[see][]{ohno_nitrogen_2023, ohno_nitrogen_2023-1}.

\subsection{1D chemical modelling}
\label{sec:1D_chem}
We examine if our retrieved dayside hot spot thermal structure in \ref{sec:thermal_structure} can produce our retrieved molecular abundances in \ref{sec:thermal_structure}, assuming our tentatively estimated metallicity and C/O ratio in \ref{sec:abundance}.
This is motivated by our assumption that the atmospheric chemical abundances are homogenised to the dayside values by circulation in the pressure region probed by our observation \citep{cooper_dynamics_2006, agundez_pseudo_2014, mendonca_three-dimensional_2018, venot_global_2020, baeyens_grid_2021}.
We employ a 1D chemical model \citep{tsai_vulcan_2017,tsai_comparative_2021}, which simultaneously takes into account thermochemistry, photochemistry and vertical mixing.
In particular, we want to explore if we can produce the level of NH$_3$ retrieved in \ref{sec:abundance}, which is more than one order of magnitude higher than several model predictions for a WASP-43b like planet (\citealt{venot_global_2020,baeyens_grid_2021}; Figure 14 and 21 of \citealt{fortney_beyond_2020}).  
We extend our retrieved representative dayside TP profile in \ref{sec:thermal_structure} down to 100 bar using the double-grey analytical expression of \cite{heng_analytical_2014} to model the chemistry in the deep atmosphere, assuming a T$_{\text{int}}$ of 595 K.
We find that the temperature of our retrieved dayside profile $T_{\text{day}}$ at 1 bar is so high that the vertical quenching of chemical abundances happens at approximately 1 bar.
Therefore, the thermal structure below the 1 bar pressure level does not impact the photospheric composition, and the results of our 1D chemical modelling is insensitive to the particular value of T$_{\text{int}}$ that we choose, since we always have to match our retrieved TP profile at around 1 bar.
We then test three different metallicities corresponding to the median and the 1$\sigma$ central credible interval bounds of our estimated metallicity while keeping the input C/O ratio at 0.8. 
We also experiment with different values of vertical eddy diffusion coefficient $K_{zz}$ \citep[e.g.,][]{parmentier_3d_2013} and N/H ratios to see if we can produce our retrieved NH$_3$ abundance using the 1D chemical model.

\begin{figure}
\centering
\includegraphics[scale=0.51]{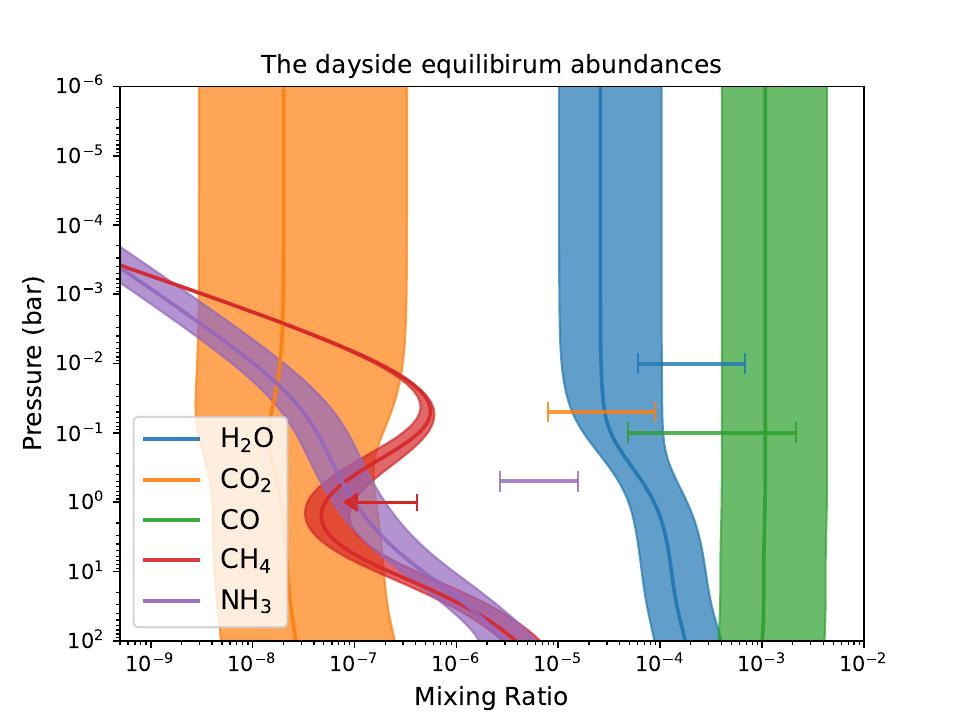}
\caption{ 
Comparison of dayside thermochemical equilibrium abundances with the chemical abundances retrieved from the MIRI data.
The shaded regions represent a range of 1D thermochemical equilibrium models consistent with our estimated metallicity in \ref{sec:abundance}.
The horizontal error bars show the 1$\sigma$ credible intervals of our retrieved molecular abundances using the MIRI data. 
The right edge  of the red horizontal arrow shows the 95\% upper limit of CH$_4$ abundance retrieved using the MIRI data.
Note that the vertical coordinates of the error bars and the arrow are arbitrary. 
}
\label{fig:dayside_eq}
\end{figure}

\begin{figure}
\centering
\includegraphics[scale=0.51]{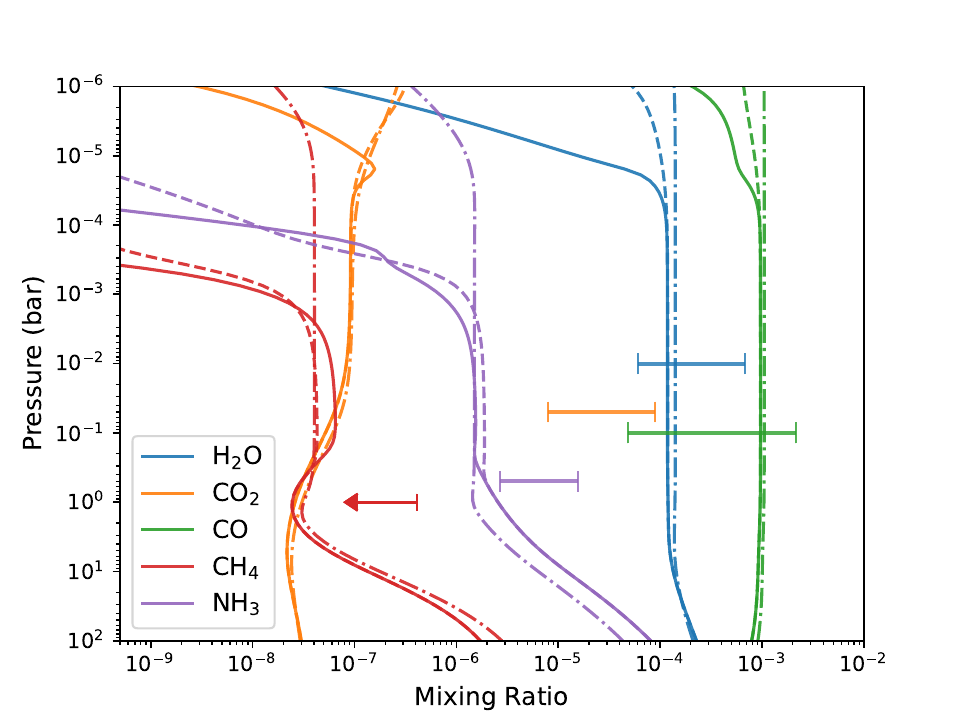}
\caption{Dayside chemical abundances with vertical mixing and photochemistry. 
The solid and dashed lines show the model with 1000$\times$ enriched N elemental abundance, with the solid lines showing $K_{zz} = 10^8$ cm$^2$/s and the dashed lines showing $K_{zz} = 10^{10}$ cm$^2$/s. 
The dash-dotted lines show the model with 200$\times$ enriched N elemental abundance and $K_{zz} = 10^{13}$ cm$^2$/s.
Note that the vertical coordinates of the error bars and the arrow are arbitrary. 
}
\label{fig:dayside_mix}
\end{figure}

We now describe the results of our 1D chemical modelling.
We first examine the dayside chemical abundances under thermochemical equilibrium without vertical mixing and photochemistry. 
As shown in Figure \ref{fig:dayside_eq}, our retrieved H$_2$O and CO abundances are consistent with their equilibrium abundances, which are roughly uniform with pressure. 
Our retrieved CH$_4$ upper bound is also consistent with the dayside equilibrium abundance, which is lower than $10^{-6}$ for pressures lower than 10 bar.  
However, our retrieved abundances for CO$_2$ and NH$_3$ are much higher than the equilibrium abundances. 
We defer the investigation of CO$_2$ as its detection significance is relatively low and focus on the issue of NH$_3$.
We include the effect of photochemistry and vertical mixing on the equilibrium abundances in Figure \ref{fig:dayside_mix}. 
We show three models that can increase the level of NH$_3$ compared to the equilibrium model: (1) 200$\times$ enhanced N/H, $K_{zz} = 10^{13}$ cm$^2$/s (dash-dotted lines); (2) 1000$\times$ enhanced N/H, $K_{zz} = 10^{8}$ cm$^2$/s (solid lines); (3) 1000$\times$ enhanced N/H, $K_{zz} = 10^{10}$ cm$^2$/s (dashed lines).
We illustrate two pathways to increasing the NH$_3$ abundance in the context of a one-dimensional model: invoking stronger vertical mixing to quench the NH$_3$ abundance to deeper pressure levels or increasing the overall N abundance in the atmosphere.
In the context of our 1D chemical modelling, an improbably strong enhancement of N/H ($>$200$\times$ enhanced N/H) is necessary to obtain our retrieved NH$_3$ abundance, even with unlikely high vertical mixing. 
The real picture, with 3D circulation, is much more complex, and we plan to investigate the atmospheric chemistry of WASP-43b more comprehensively in a future paper.

\section{Limitations}
\label{sec:assumptions}
Given the constraints in data quality, in particular the lack of spatial resolution, and the gaps in our theoretical understanding of hot Jupiter atmospheres, we have to make a series of assumptions in our modelling. 
We now discuss the validity of our assumptions and how they impact the conclusions we draw from the data.

\subsection{Thermal structure}
\label{sec:thermal_assumption}
Our 2D parametric temperature model comes with various assumptions.
We first assume that a latitudinally-averaged structure can effectively model the phase curve data. 
In our 2D temperature model, temperature varies as a function of longitude and pressure while being held constant as a function of latitude on isobars (see equation (\ref{eq:Tmap}) and Figure \ref{fig:model}). 
We then interpret the retrieved temperature map as a meridional mean temperature map, using cos(latitude) as the weight.
We base this interpretation on the fact that \cite{irwin_25d_2020} and \cite{yang_testing_2023} find the temperature structure retrieved from synthetic phase curve data simulated from a GCM closely resembles the latitudinally-averaged thermal structure of the GCM using cos(latitude) as the weight. 
Our temperature model further assumes that the thermal structure is symmetric about the equator (north-south symmetry) and about the meridian going through the centre of the dayside region (east-west symmetry) and that the thermal structure on the nightside is more homogeneous than that on the dayside. 
As discussed in \ref{sec:model}, this meridional mean structure is inspired by GCM studies, though we cannot explore some of the more complex mean thermal structures seen in GCM simulations, such as pressure-dependent hot spot offset.
Our parametric modelling is only one of the first steps in understanding the thermal structure of hot Jupiters through observations, and we expect that as more \textit{JWST} data come in, a more sophisticated model set-up will be necessary.
Looking ahead, we can constrain the dayside latitudinal thermal variation by using eclipse mapping data \citep{ rauscher_toward_2007,challener_eclipse-mapping_2023, boone_analytical_2024}, where high cadence flux measurements are taken as the star gradually eclipses the planet. 
In future work, we plan to incorporate MIRI eclipse mapping data of WASP-43b \citep{hammond_two-dimensional_2024} in retrievals to constrain the 3D thermal structure of WASP-43b.

\subsection{Clouds}
We assume that the atmosphere of WASP-43b is free from significant aerosols in the pressure range probed by the observation.
The assumption that the dayside region contains no optically thick aerosols in the pressure ranges probed by MIRI is supported by numerous studies: transmission spectroscopy with \textit{HST}/WFC3 \citep{kreidberg_precise_2014}, inference of bond albedo from \textit{HST}/WFC3 + \textit{Spitzer}/IRAC phase curves \citep{stevenson_spitzer_2017}, reflected light photometry with \textit{HST} WFC3/UVIS \citep{fraine_dark_2021}, a joint analysis of \textit{CHEOPS}, \textit{TESS}, and \textit{HST} WFC3/UVIS data \citep{scandariato_phase_2022}, and a retrieval study of VLT/CRIRES+ high-resolution dayside emission spectra \citep{lesjak_retrieval_2023}.
Furthermore, a cloud-free dayside photosphere is consistent with the high dayside temperatures we retrieve in this work.
On the other hand, a cloudy nightside has been inferred from large day-night brightness temperature contrast \citep{stevenson_spitzer_2017,kataria_atmospheric_2015,mendonca_revisiting_2018,morello_independent_2019-1, irwin_25d_2020}. 
\cite{bell_nightside_2024} also find that cloudy GCMs fit the phase curve at nightside phases better than cloudless GCMs. 
Notably, both the retrieval analysis in \cite{bell_nightside_2024} and the retrieval analysis in this work show that the observed spectra can be adequately fit with cloud-free parametric atmospheric models. 
As discussed in \ref{sec:thermal_structure}, it has been shown that a more isothermal temperature profile can mimic the effects of cloud opacity in low-resolution spectral retrievals \citep{burningham_retrieval_2017, molliere_retrieving_2020}, which explains the ability of simple models to fit the observation without invoking cloud opacity.
The omission of clouds from our model means that our retrieved nightside thermal structure would not be reliable if the nightside of WASP-43b is, in fact, cloudy.
However, if our assumption that the atmospheric chemical abundances are constant with longitude and pressure holds, then the likely cloud-free dayside still anchors our retrieved chemical abundances.
Since we retrieve spectra from four different orbital phases simultaneously, we do not expect our retrieved chemical abundances to be significantly biased.
On the other hand, as mentioned in \ref{sec:abundance}, even though our abundance constraints may not be biased in the presence of nightside clouds, a significant amount of oxygen may be removed from the gas-phase molecules if the cloud composition is rich in silicates \citep{woitke_equilibrium_2018}.
Consequently, based on our gas-phase abundance constraints, we would underestimate the metallicity and overestimate the C/O ratio of WASP-43b.

\subsection{Chemistry}
\label{sec:chemistry}
We assume that the VMRs of all molecules in our model atmosphere are constant with longitude, latitude, and pressure.
This assumption holds well for H$_2$O and CO, as both thermochemical equilibrium models and models that include photochemistry and mixing predict H$_2$O and CO abundances to be approximately uniform in the observed pressure region for a WASP-43b like planet \cite[e.g., Figure 5,][]{venot_global_2020}. 
The assumption of uniform H$_2$O abundance is supported by the retrieval analysis in \cite{bell_nightside_2024}, which showed no significant phase variation in H$_2$O abundance (see Figure \ref{fig:compare_bell24}).
On the other hand, the spatial distributions of CH$_4$, NH$_3$ and CO$_2$ in thermochemical equilibrium vary significantly with longitude and pressure, especially in the case of CH$_4$, which would be present at a detectable level in the nightside of WASP-43b under thermochemical equilibrium.
However, by coupling a simplified chemical kinetics scheme to a GCM of HD 209458b, \cite{cooper_dynamics_2006} showed that the CH$_4$ abundance should be homogenised between 1 bar and 1 millibar to be representative of the low dayside abundance when strong zonal circulation is present.
The uniform CH$_4$ prediction was later corroborated by \cite{agundez_pseudo_2014}, who coupled a more realistic chemical network to a simplified circulation scheme that included both vertical mixing and a uniform zonal wind.
Modelling specific to the planetary parameters of WASP-43b showed that dynamical timescale dominates over chemical timescale between 1 bar and 1 millibar \citep{mendonca_three-dimensional_2018, venot_global_2020}, and we expect no significant longitudinal variation in the abundances of CO$_2$, CH$_4$ and NH$_3$.  
The assumption of constant-with-longitude CH$_4$ abundance is supported by the non-detection of CH$_4$ at all orbital phases in \cite{bell_nightside_2024}. 
There is also circumstantial evidence of strong zonal circulation on WASP-43b, for example, the $\sim$8 degrees east hot spot shift that is retrieved in this work and in \cite{bell_nightside_2024}, and a wind speed of 3-8 kms$^{-1}$ inferred from high-resolution emission spectroscopy \citep{lesjak_retrieval_2023}.
While theoretical models do predict some vertical variation in the abundances of  CO$_2$, CH$_4$ and NH$_3$, the variation in CO$_2$ with pressure is within one order of magnitude and thus likely insignificant compared to the uncertainties in its retrieved abundance. 
The vertical variations of CH$_4$ and NH$_3$ warrant further investigation and a parameterisation of vertical quenching may be required if we are sensitive to a wide range of pressures. 
However, as shown in Figure \ref{fig:twf_day} and Figure \ref{fig:twf_night}, the transmission weighting function of our best-fit model peaks strongly between 1 bar and 0.1 bar.
Given the limited pressure range probed by our observation, we feel the modelling assumption that the abundances of CH$_4$ and NH$_3$ are constant with pressure is justified.

We assume a H$_2$/He dominated model atmosphere and only include five spectrally active molecules: H$_2$O, CO$_2$, CO, CH$_4$, and NH$_3$.
We chose this set of molecules based on chemical modelling of hot Jupiter atmospheres \citep{agundez_pseudo_2014, venot_global_2020,baeyens_grid_2021}, taking into account the opacities of molecules, as well as guided by previous retrieval studies of WASP-43b such as \cite{bell_nightside_2024} and \cite{yang_testing_2023}.
While this set of molecules adequately explains the spectra, we could not fit an absorption signature at 8.75 $\micron$ that only appears at phase 0.5.
The investigation of more chemical species and spatial variation in chemistry is left for future work.

\subsection{Radiative transfer}
We assume that the correlated-k approximation for radiative transfer calculations \citep{lacis_description_1991} introduces negligible errors in modelling exoplanet spectra, as found by \cite{molliere_model_2015}, \cite{irwin_25d_2020}, and \cite{leconte_spectral_2021}. 
Furthermore, we use channel-averaged k-tables binned to our data's 0.5 $\micron$ spectral resolution. 
We assess the accuracy of this approximation by computing spectra from our best-fit atmospheric model using k-tables at a resolution of $\text{R}=1000$.
The results are in good agreement with the spectra calculated using k-tables at the spectral resolution of our data: the differences between the two spectra are found to be well within the measurement uncertainties.
The errors introduced by our radiative transfer method are thus unlikely to bias the retrieved physical parameters significantly.
Since our Bayesian inference algorithm requires the generation of hundreds of thousands of models to explore the model parameter space, a faster radiative transfer routine is much preferred if there is no significant loss in accuracy.
Nonetheless, as data quality advances, we need to more carefully assess the errors introduced in our forward modelling, particularly our radiative transfer calculation method, as well as the errors intrinsic to the opacity data we employ.

\subsection{Data reduction}
In this work, we are treating the spectra at different orbital phases as independent measurements to boost the signal-to-noise ratio of the retrieval. 
In reality, these spectra are correlated through the data reduction process, as phase curves are conventionally extracted by fitting a combination of Fourier series and systematics model to the time-series observations, wavelength by wavelength \citep{bell_nightside_2024, changeat_toward_2024}.
In effect, we have assumed that the measurement uncertainties of our spectra are Gaussian and independent, when in reality they are non-Gaussian and correlated. 
Consequently, we expect our posterior distribution to be biased.
Given the goodness-of-fit of our model to the data in Figure \ref{fig:spectral_fit}, we expect our retrieved molecular abundances are still consistent with the data.
A full account of this issue is beyond the scope of this study, and we refer the reader to a possible resolution by \cite{changeat_toward_2024}, where the atmospheric retrieval is performed simultaneously with data reduction on the time-series observations.

\section{Conclusion}
\label{sec:conclusion}
We find evidence of H$_2$O (6.5$\sigma$), NH$_3$ ($4\sigma$), CO (3.1$\sigma$), CO$_2$ (2.5$\sigma$) and no evidence of CH$_4$ in the atmosphere of the hot Jupiter WASP-43b by retrieving four emission spectra observed at different orbital phases using the MIRI/LRS on the \textit{JWST}. 
We fit the spectra simultaneously by employing a 2D atmospheric model accounting for a non-uniform temperature profile across the dayside and assuming uniform chemical abundances \citep{yang_testing_2023}.
Our approach more confidently and precisely constrains the molecular abundances than the previous retrievals in \cite{bell_nightside_2024}, which analysed each orbital phase separately.
Based on our abundance constraints, we tentatively estimate the metallicity of WASP-43b at $1.6^{+4.9}_{-1.0}\times$ solar and its C/O ratio at $0.8^{+0.1}_{-0.2}$, though we stress the need for further observations as the detection significance of the carbon-bearing molecules in this study are relatively low.
In particular, the scheduled \textit{JWST} NIRSpec observation (GTO 1224, PI: S. Birkmann) will likely revise these estimates by better constraining the abundances of CO and CO$_2$, giving us an even clearer picture of the bulk chemistry of WASP-43b.
Our tentative detection of NH$_3$, seen for the first time in the emission spectra of a hot Jupiter, may be verified in the future by reliably reducing the 10.6-12 $\micron$ region of our MIRI/LRS data that is excluded from our retrieval due to residual instrument systematics, or by further observations such as the scheduled \textit{JWST} NIRSpec observation.
The precise abundance constraints in this work showcase the need for multidimensional atmospheric models to retrieve phase curve data and open up exciting avenues for follow-up studies on the atmospheric chemistry and formation history of WASP-43b.

\section*{Acknowledgements}
This work is based on observations made with the NASA/ESA/CSA JWST. The data were obtained from the Mikulski Archive for Space Telescopes at the Space Telescope Science Institute, which is operated by the Association of Universities for Research in Astronomy, Inc., under NASA contract NAS 5-03127 for \textit{JWST}. These observations are associated with program JWST-ERS-01366. Support for program JWST-ERS-01366 was provided by NASA through a grant from the Space Telescope Science Institute.

We thank Adina Feinstein, Luis Welbanks and Michael Line for developing the software to plot Figure 4.
We thank Patricio Cubillos for providing the routine to plot Figure 9.
JY thanks Olivia Gallup for proofreading the manuscript.
JY acknowledges Alex Cridland, Luke Parker and Joost Wardenier for their helpful discussions.
MDH acknowledges funding from Christ Church, University of Oxford. 
TJB acknowledges funding support from the NASA Next Generation Space Telescope Flight Investigations program (now \textit{JWST}) via WBS 411672.07.05.05.03.02. 
JKB is supported by an STFC Ernest Rutherford Fellowship, grant number ST/T004479/1.

\section*{Data availability}
The data used in this paper are associated with \textit{JWST} DD-ERS program 1366 (PIs Batalha, Bean, and Stevenson; observation 11) and are publicly available from the Mikulski Archive for Space Telescopes (\url{https://mast.stsci.edu}).
Additional intermediate and final results from \cite{bell_nightside_2024} that are analysed in this work will be archived on Zenodo at
\url{https://zenodo.org/doi/10.5281/zenodo.10525170} upon final publication of \cite{bell_nightside_2024}.
All other data underlying this article will be shared on reasonable request to the corresponding author.




\bibliographystyle{mnras}
\bibliography{miri} 



\appendix
\section{Full posterior}
\label{appendix:full_posterior}
\begin{figure*}
\includegraphics[scale=0.18]{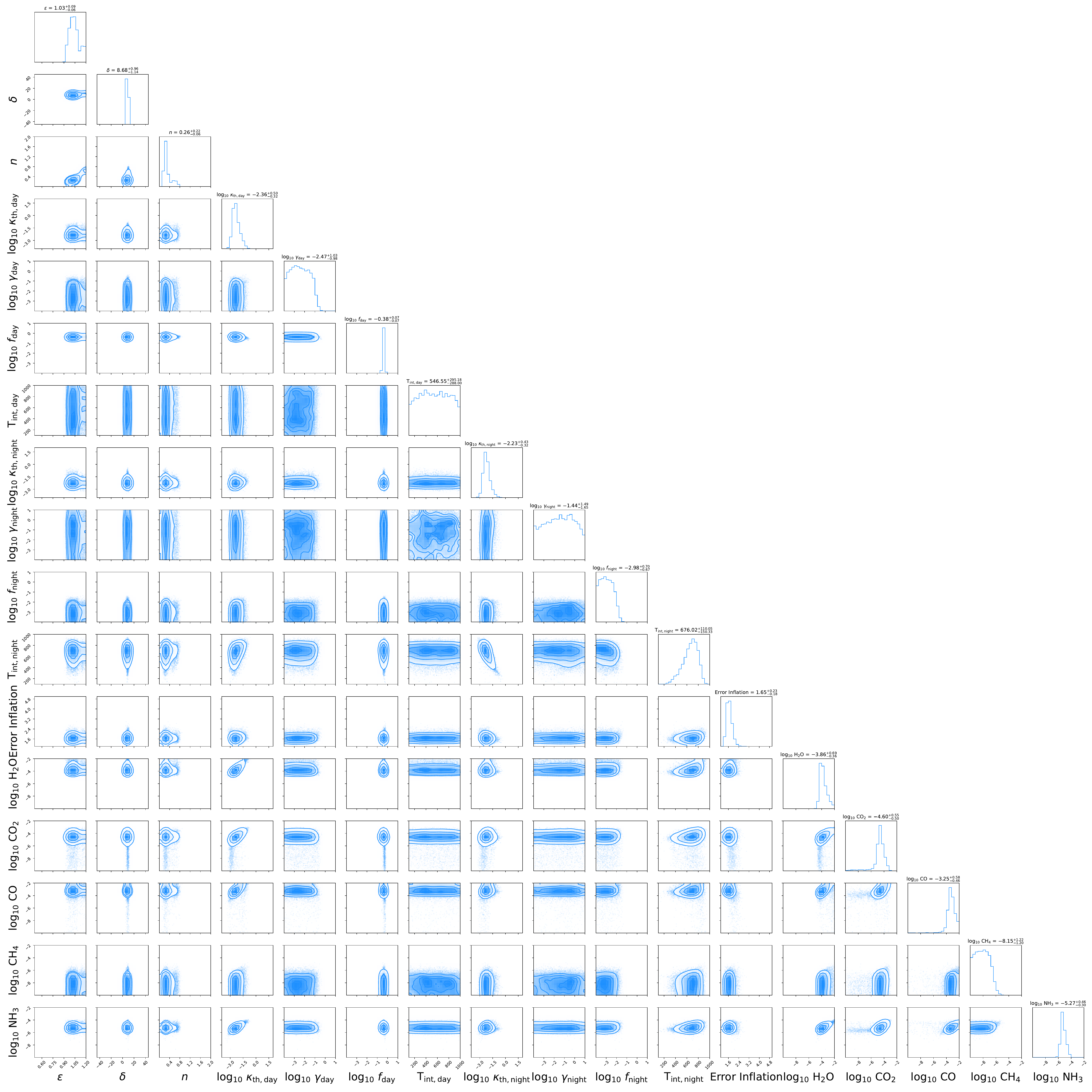}
\caption{Posterior distribution of all the model parameters listed in Table \ref{tab:priors}.
The diagonal plots are the marginalised posterior distributions, and the labels above show the posterior medians and 1$\sigma$ central credible intervals.
}
\label{fig:full_posterior}
\end{figure*}

Figure \ref{fig:full_posterior} shows the full posterior distribution of the retrieval model parameters listed in Table \ref{tab:priors}. 

\section{Comparison with past studies}
\label{appendix:comparison}
We compare our retrieved H$_2$O abundance to past studies that retrieved H$_2$O abundance.
Note that apart from \cite{lesjak_retrieval_2023}, who analysed ground-based high-resolution spectra observed with VLT/CRIRES+, all other studies analysed low-resolution space telescope data.
For \cite{bell_nightside_2024}, we list the  phase-by-phase retrieval results from the three retrieval pipelines that included H$_2$O as a free parameter (see also Figure \ref{fig:compare_bell24} in the main text). 

\begin{figure*}
\includegraphics[scale=.72]{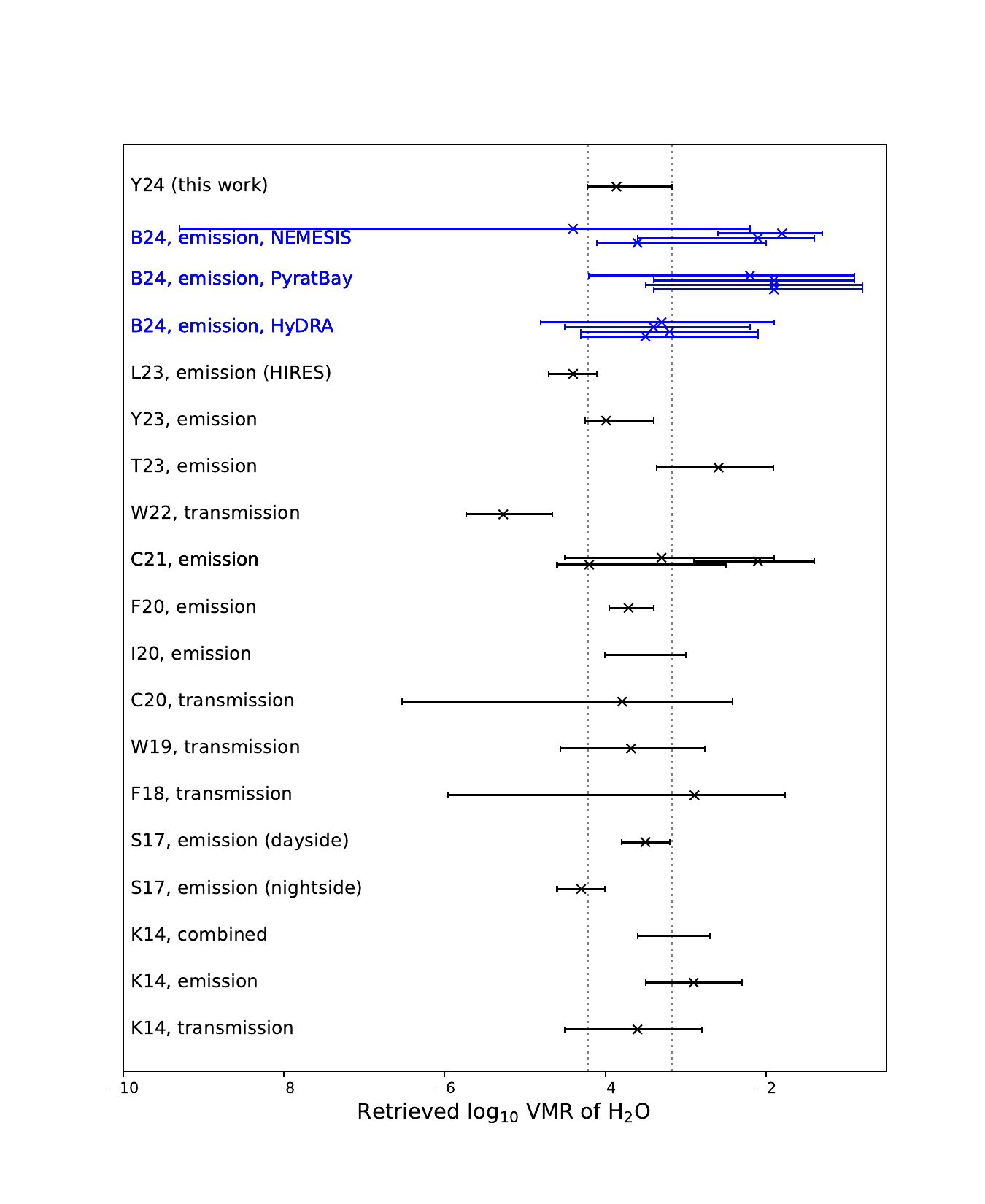}
\caption{Comparison to past studies of WASP-43b that retrieved H$_2$O abundance. 
The error bars represent the 1$\sigma$ credible intervals, and the crosses mark the posterior medians for studies that quoted such values.
The vertical dashed lines mark the 1$\sigma$ credible interval retrieved in this work.
Note that both C21 and B24 performed phase-by-phase retrievals, hence we plot multiple error bars to show the constraints from all phases.
The constraints for T23 are taken from the Scat.Cloud model in their Table A6.
The list of studies is as follows:
K14: \protect\cite{kreidberg_precise_2014}; 
S17: \protect\cite{stevenson_spitzer_2017}; 
F18: \protect\cite{fisher_retrieval_2018}; 
W19: \protect\cite{welbanks_massmetallicity_2019}; 
C20: \protect\cite{chubb_aluminium_2020};
I20: \protect\cite{irwin_25d_2020}; 
F20: \protect\cite{feng_2d_2020}; 
C21: \protect\cite{cubillos_longitudinally_2021}; 
W22: \protect\cite{welbanks_atmospheric_2022}; 
T23: \protect\cite{taylor_another_2023}; 
Y23: \protect\cite{yang_testing_2023}; 
L23: \protect\cite{lesjak_retrieval_2023}; 
B24: \protect\cite{bell_nightside_2024}; 
Y24: this work.
Note that K14, S17, F18, W19, C20, I20, F20, C21, W22, T23, Y23 analysed subsets of \textit{HST}/WFC3 and \textit{Spitzer}/IRAC data, L23 analysed VLT/CRIRES+ data, and B24 and Y24 analysed \textit{JWST}/MIRI data.
}
\label{fig:compare_all}
\end{figure*}



\bsp	
\label{lastpage}
\end{document}